\documentclass[aps,prl,reprint,superscriptaddress]{revtex4-2}

\usepackage{graphicx}
\usepackage{physics}
\usepackage{xcolor}
\usepackage{float}
\usepackage[section]{placeins}
\usepackage{chapterbib}

\def\CRO{Ca$_2$RuO$_4$}

\begin{document}

	\title{Keldysh space control of charge dynamics in a strongly driven Mott insulator}
	
	\normalsize
	
	\author{Xinwei Li}
	\thanks{These authors contributed equally to this work}
	\affiliation{Institute for Quantum Information and Matter, California Institute of Technology, Pasadena, CA 91125, USA}
	\affiliation{Department of Physics, California Institute of Technology, Pasadena, CA 91125, USA}
	
	\author{Honglie Ning}
	\thanks{These authors contributed equally to this work}
	\affiliation{Institute for Quantum Information and Matter, California Institute of Technology, Pasadena, CA 91125, USA}
	\affiliation{Department of Physics, California Institute of Technology, Pasadena, CA 91125, USA}
	
	\author{Omar Mehio}
	\thanks{These authors contributed equally to this work}
	\affiliation{Institute for Quantum Information and Matter, California Institute of Technology, Pasadena, CA 91125, USA}
	\affiliation{Department of Physics, California Institute of Technology, Pasadena, CA 91125, USA}

	\author{Hengdi Zhao}
	\affiliation{Department of Physics, University of Colorado, Boulder, CO 80309, USA}
	
	\author{Min-Cheol Lee}
	\affiliation{Center for Integrated Nanotechnologies, Los Alamos National Laboratory, Los Alamos, NM 87545, USA}
	\affiliation{Center for Correlated Electron Systems, Institute for Basic Science, Seoul 08826, Republic of Korea}
	
	\author{Kyungwan Kim}
	\affiliation{Department of Physics, Chungbuk National University, Cheongju, Chungbuk 28644, Republic of Korea}
	
	\author{Fumihiko Nakamura}
	\affiliation{Department of Education and Creation Engineering, Kurume Institute of Technology, Fukuoka 830-0052, Japan}
	
	\author{Yoshiteru Maeno}
	\affiliation{Department of Physics, Graduate School of Science, Kyoto University, Kyoto 606-8502, Japan}
	
	\author{Gang Cao}
	\affiliation{Department of Physics, University of Colorado, Boulder, CO 80309, USA}
	
	\author{David Hsieh}
	\email[email: ]{dhsieh@caltech.edu}
	\affiliation{Institute for Quantum Information and Matter, California Institute of Technology, Pasadena, CA 91125, USA}
	\affiliation{Department of Physics, California Institute of Technology, Pasadena, CA 91125, USA}
	
	\date{\today}
	
	\begin{abstract}
		The fate of a Mott insulator under strong low frequency optical driving conditions is a fundamental problem in quantum many-body dynamics. Using ultrafast broadband optical spectroscopy, we measured the transient electronic structure and charge dynamics of an off-resonantly pumped Mott insulator Ca$_2$RuO$_4$. We observe coherent bandwidth renormalization and nonlinear doublon-holon pair production occurring in rapid succession within a sub-100 femtosecond pump pulse duration. By sweeping the electric field amplitude, we demonstrate continuous bandwidth tuning and a Keldysh cross-over from a multi-photon absorption to quantum tunneling dominated pair production regime. Our results provide a procedure to control coherent and nonlinear heating processes in Mott insulators, facilitating the discovery of novel out-of-equilibrium phenomena in strongly correlated systems.
	\end{abstract}

	\maketitle

	The response of a Mott insulator to a strong electric field is a fundamental question in the study of non-equilibrium correlated many-body systems \cite{Eckstein2010,Lee2014,Lenarseci2012,Li2015,Oka2005,Diener2018,Asamitsu1997,Chu2020,Wall2011,Okamoto2007,Mitrano2014,Strohmaier2010,Lenarseci2013,Sensarma2010,Eckstein2011}. In the DC limit, a breakdown of the insulating state occurs when the field strength exceeds the threshold for producing pairs of doubly-occupied (doublon) and empty (holon) sites by quantum tunneling, in analogy to the Schwinger mechanism for electron-positron pair production out of the vacuum \cite{Schwinger1951}. Recently, the application of strong low frequency AC electric fields has emerged as a potential pathway to induce insulator-to-metal transitions \cite{Murakami2018,Giorgianni2019,Mayer2015,Yamakawa2017}, realize efficient high-harmonic generation \cite{Imai2020,Silva2018}, and coherently manipulate band structure and magnetic exchange interactions in Mott insulators \cite{Mentink2015,Hejazi2019,Mikhaylovskiy2015,Batignani2015,Claassen2017,Wang2018}. Therefore there is growing interest to understand doublon-holon (d-h) pair production and their non-thermal dynamics in the strong field AC regime.  
	
	Strong AC field induced d-h pair production has been theoretically studied using Landau-Dykhne adiabatic perturbation theory \cite{Oka2012} along with a suite of non-equilibrium numerical techniques \cite{Murakami2018,Tsuji2011,Oka2012,Imai2020,Silva2018,Takahashi2008,TancogneDejean2020}. Notably, d-h pairs are primarily produced through two nonlinear mechanisms: multi-photon absorption and quantum tunneling \cite{Oka2012,Kruchinin2018}. The two regimes are characterized by distinct electric field scaling laws and momentum space distributions of d-h pairs. By tuning the Keldysh adiabaticity parameter $\gamma_\text{K}=\hbar\omega_\text{pump}/(eE_\text{pump}\xi)$ through unity, where $\omega_\text{pump}$ is the pump frequency, $E_\text{pump}$ is the pump electric field, $e$ is electron charge, and $\xi$ is the d-h correlation length, a cross-over from a multi-photon dominated ($\gamma_\text{K}>1$) to a tunneling dominated ($\gamma_\text{K}<1$) regime can in principle be induced. However, direct experimental tests are lacking owing to the challenging need to combine strong tunable low frequency pumping fields with sensitive ultrafast probes of non-equilibrium distribution functions.
	
	We devise a protocol to study these predicted phenomena using ultrafast broadband optical spectroscopy. As a testbed, we selected the multiband Mott insulator \CRO. Below a metal-to-insulator transition temperature $T_\text{MIT}=357$~K, a Mott gap ($\Delta=0.6$~eV) opens within its 2/3-filled Ru 4$d$ $t_{2g}$ manifold \cite{Gorelov2010,Han2018,Fang2004,Jung2003}, with a concomitant distortion of the lattice \cite{Braden1998}. Upon further cooling, the material undergoes an antiferromagnetic transition at $T_\text{N}$ = 113 K into a N\'{e}el ordered state. It has recently been shown that for temperatures below $T_\text{MIT}$, re-entry into a metallic phase can be induced by a remarkably weak DC electric field of order 100 V/cm \cite{Nakamura2013}, making Ca$_2$RuO$_4$ a promising candidate for exhibiting efficient nonlinear pair production. 
	
	\begin{figure*}[t!]
		\centering
		\includegraphics[width=0.85\linewidth]{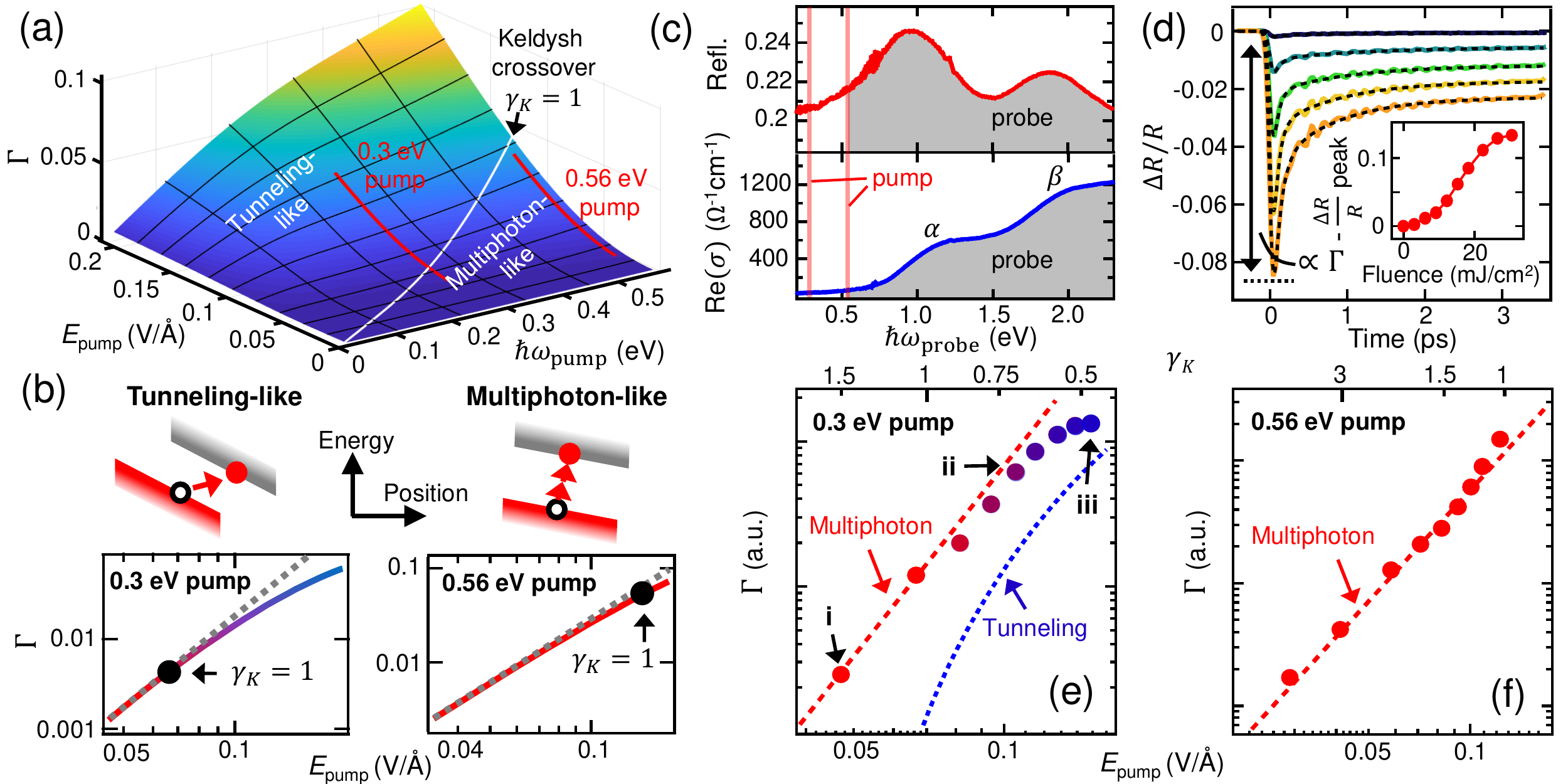}
		\caption{\small Resolving Keldysh tuning using pump-probe spectroscopy. (a)~$\Gamma$ calculated across Keldysh space using the Landau-Dykhne method. (b)~Constant energy cuts along the red lines shown in (a) plotted on a logarithmic scale. Black dots mark the Keldysh cross-over. Gray dashed lines: scaling relation in the multi-photon regime. Schematics of the multi-photon and tunneling processes are shown above. (c)~Equilibrium reflectivity (top) and conductivity (bottom) spectra of \CRO\ at 20~K. The 0.3~eV and 0.56~eV pump energies are marked by vertical red lines. The probe energy range is shaded grey. (d)~Select 0.3~eV pump 1.77~eV probe $\Delta R/R$ traces at fluences of $3$, $9$, $15$, $22$, and $30$~mJ/cm$^2$ (top to bottom). Dashed lines are fits detailed in \cite{SM}. Inset: Peak $\Delta R/R$ versus fluence showing nonlinearity. (e, f) Experimental cuts through the same regions of parameter space as in (b). Error bars are smaller than data markers. Scaling relations for multi-photon and tunneling behavior are overlaid as red and blue dashed lines respectively.
		}
		\label{Fig1}
	\end{figure*}
	
	To estimate the response of Ca$_2$RuO$_4$ to a low frequency AC electric field, we calculated the d-h pair production rate ($\Gamma$) over the Keldysh parameter space using a Landau-Dykhne method developed by Oka \cite{Oka2012}. Experimentally determined values of the Hubbard model parameters for Ca$_2$RuO$_4$ were used as inputs \cite{SM}. As shown in Figure~1(a), $\Gamma$ is a generally increasing function of $E_\text{pump}$ and $\hbar\omega_\text{pump}$. For a fixed $\omega_\text{pump}$, the predicted scaling of $\Gamma$ with $E_\text{pump}$ is clearly different on either side of the Keldysh cross-over line ($\gamma_\text{K}=1$), evolving from power law behavior $\Gamma\propto(E_\text{pump})^a$ in the multi-photon regime to threshold behavior $\Gamma\propto\text{exp}(-b/E_\text{pump})$ in the tunneling regime [Fig. 1(b)]. 
	
	At time delays where coherent nonlinear processes are absent, the transient pump-induced change in reflectivity of a general gapped material is proportional to the density of photo-excited quasi-particles \cite{Gedik2004,Chia2006,Demsar1999}, which, upon dividing by a constant pump pulse duration ($\sim$100~fs), yields $\Gamma$. Differential reflectivity ($\Delta R/R$) transients from Ca$_2$RuO$_4$ single crystals were measured at $T$ = 80 K using several different subgap pump photon energies ($\hbar\omega_\text{pump} < \Delta$) in the mid-infrared region, and across an extensive range of probe photon energies ($\hbar\omega_\text{probe}$) in the near-infrared region spanning both the $\alpha$ and $\beta$ absorption peaks [Fig. 1(c)]. These two band edge features can be assigned to optical transitions within the Ru $t_{2g}$ manifold \cite{Das2018,Jung2003}. Figure 1(d) shows reflectivity transients at various fluences measured using $\hbar\omega_\text{pump}$ = 0.3~eV and $\hbar\omega_\text{probe}$ = 1.77~eV. Upon pump excitation, we observe a rapid resolution-limited drop in $\Delta R/R$. With increasing fluence, the minimum value of $\Delta R/R$ becomes larger, indicating a higher value of $\Gamma$ within the pump pulse duration. This is followed by exponential recovery as the d-h pairs thermalize and recombine \cite{SM}. By plotting $\Gamma$ against the peak value of $E_\text{pump}$ (measured in vacuum), we observe a change from power law scaling to threshold behavior when $E_\text{pump}>0.07$~V/\AA~[Fig. 1(e)], in remarkable agreement with our calculated Keldysh cross-over [Figs. 1(a),(b)]. In contrast, measurements performed using 0.56 eV pumping exhibit exclusively power law scaling over the same $E_\text{pump}$ range [Fig. 1(f)], again consistent with our model.  
	
	\begin{figure*}[t!]
		\centering
		\includegraphics[width=0.95\linewidth]{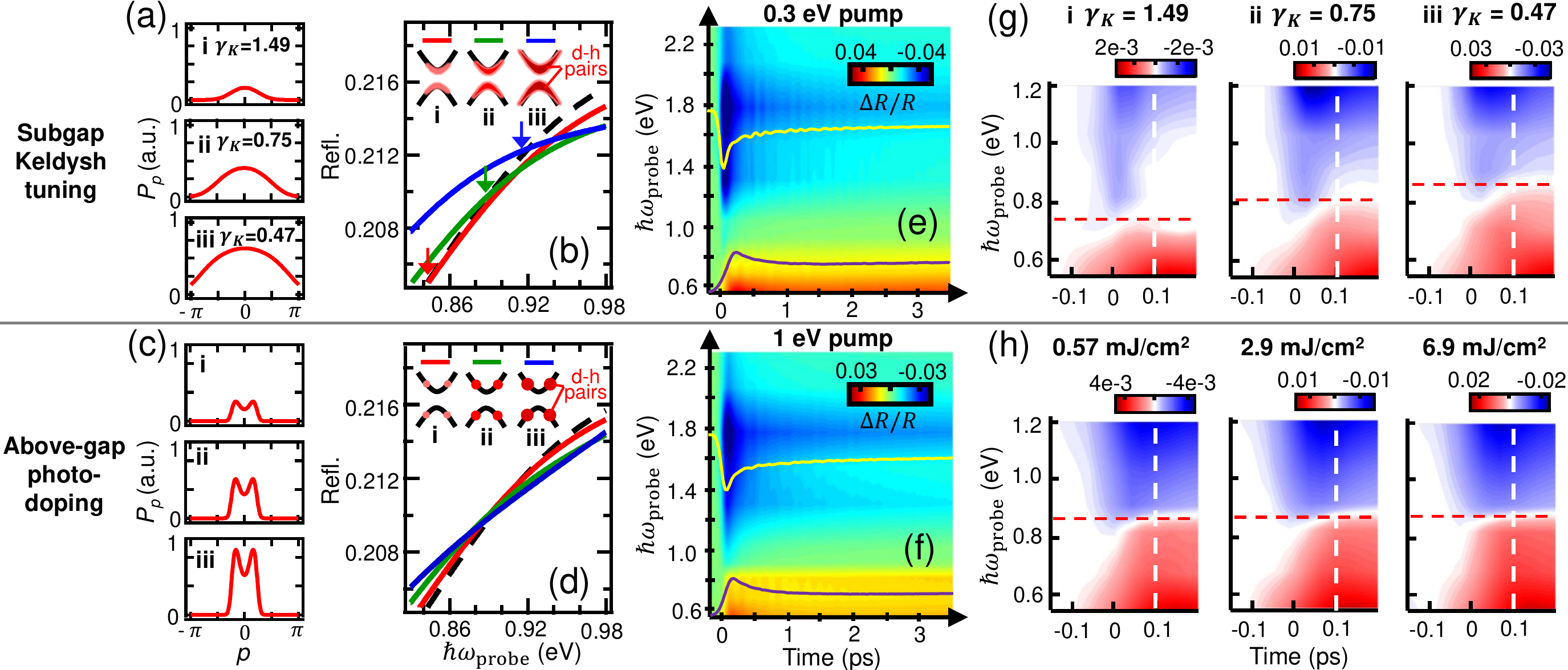}
		\caption{\small Non-thermal pair distribution through the Keldysh crossover. (a)~Calculated $P_p$ for conditions \textbf{i} to \textbf{iii} using the Landau-Dykhne method. (b)~Simulated non-equilibrium reflectivity spectra for subgap pumping. (c, d)~ Analogues of (a) and (b) but simulated for above-gap pumping. Fluence increases from \textbf{i} to \textbf{iii}. Black curves in (b) and (d) are the equilibrium spectra. Arrows in (b) mark the crossing points between the non-equilibrium and equilibrium curves.  Experimental $\Delta R/R$ maps of Ca$_2$RuO$_4$ for (e) 0.3~eV pump (fluence: 30~mJ/cm$^2$) and (f) 1~eV pump (fluence: 7~mJ/cm$^2$). Two representative constant energy cuts (yellow: 1.77~eV, purple: 0.56~eV) are overlaid. (g)~Enlargement of $\Delta R/R$ maps for 0.3~eV pump using three pump fluences [marked in Fig.\,1(e)] corresponding to conditions (i) to (iii) in (a). (h)~~Enlargement of $\Delta R/R$ maps for 1~eV pump using three pump fluences indicated above. White dashed lines mark $t=0.1$~ps. Red dashed lines: guides to the eye for the $\hbar\omega_\text{probe}$ where $\Delta R/R$ changes sign at $t=0.1$~ps. 
		}
		\label{Fig2}
	\end{figure*}
	
	A predicted hallmark of the Keldysh cross-over is a change in width of the non-thermal distribution of d-h pairs in momentum space \cite{Oka2012}. In the multi-photon regime, doublons and holons primarily occupy the conduction and valence band edges respectively, resulting in a pair distribution function ($P_p$) sharply peaked about zero momentum ($p = 0$). In the tunneling regime, the peak drastically broadens, reflecting the increased spatial localization of d-h pairs. Using the Landau-Dykhne method \cite{SM}, we calculated the evolution of $P_p$ for Ca$_2$RuO$_4$ as a function of $E_\text{pump}$ through the Keldysh cross-over. Figure 2(a) displays $P_p$ curves at three successively larger $E_\text{pump}$ values corresponding to (i) $\gamma_\text{K} = 1.49$, (ii) $\gamma_\text{K} = 0.75$ and (iii) $\gamma_\text{K} = 0.47$, which show a clearly broadening width along with increasing amplitude. 
	
	To demonstrate how signatures of a changing $P_p$ width are borne out in experiments, we simulate the effects of different non-thermal electronic distribution functions on the broadband optical response of a model insulator. Assuming a direct-gap quasi-two-dimensional insulator with cosine band dispersion in the momentum plane ($p_x$, $p_y$), the optical susceptibility computed using the density matrix formalism can be expressed as \cite{rosencher,SM}: 
	
	
	\begin{equation*}
		\chi=\sum_{p_x, p_y}C\mathcal{L}[\hbar\omega_\text{probe}-\Delta(p_x, p_y)][N_v(p_x, p_y)-N_c(p_x, p_y)]
		\label{optsusc}
	\end{equation*}
	
	\noindent where $C$ is a constant incorporating the transition matrix element, $\mathcal{L}$ represents a Lorentzian oscillator centered at the gap energy $\Delta(p_x, p_y)$, and $N_v$ and $N_c$ are the occupations of the valence and conduction bands, respectively. As will be shown later [Fig. 3(a)], it is valid to assume that $\Delta(p_x, p_y)$ decreases in proportion to the number of excitations \cite{SM}. Figure 2(b) shows simulated reflectivity spectra around the band edge - converted from $\chi$ via the Fresnel equations - using Gaussian functions for $N_v$ and $N_c$ of variable width to approximate the $P_p$ lineshapes [Fig. 2(a)] \cite{SM}. As $P_p$ evolves from condition (i) to (iii), we find that the intersection between the non-equilibrium and equilibrium reflectivity spectra shifts to progressively higher energy. For comparison, we also performed simulations under resonant photo-doping conditions using the direct-gap insulator model. Figure 2(c) displays three $P_p$ curves at successively larger $E_\text{pump}$ values, which were chosen such that the total number of excitations match those in Figure 2(a). Each curve exhibits maxima at non-zero momenta where $\hbar\omega_\text{pump} = \Delta(|p|)$ is satisfied. In stark contrast to the subgap pumping case, the amplitude of $P_p$ increases with $E_\text{pump}$ but the width remains unchanged. This results in the non-equilibrium reflectivity spectra all intersecting the equilibrium spectrum at the same energy, forming an isosbestic point [Fig. 2(d)]. The presence or absence of an isosbestic point is therefore a key distinguishing feature between Keldysh space tuning and photo-doping. This criterion can be derived from a more general analytical model \cite{SM}, which shows that a key condition for identifying a Keldysh crossover is that $\Delta R/R$ spectra at difference fluences do not scale.
	
	Probe photon energy-resolved $\Delta R/R$ maps of Ca$_2$RuO$_4$ were measured in both the Keldysh tuning ($\hbar\omega_\text{pump}$ = 0.3 eV) and photo-doping ($\hbar\omega_\text{pump}$ = 1 eV) regimes. As shown in Figures 2(e) \& (f), the extremum in $\Delta R/R$, denoting the peak d-h density, occurs near a time $t$ = 0.1 ps measured with respect to when the pump and probe pulses are exactly overlapped ($t$ = 0). This is followed by a rapid thermalization of d-h pairs as indicated by the fast exponential relaxation in $\Delta R/R$, which will be discussed later \cite{SM}. Figure 2(g) shows $\Delta R/R$ maps acquired in the subgap pumping regime for three different pump fluences corresponding to conditions (i) to (iii) in Figure 2(a) and 1(e). Focusing on the narrow time window around $t$ = 0.1 ps where the d-h distribution is highly non-thermal, we observe that $\Delta R/R$ changes sign across a well-defined probe energy (dashed red line), marking a crossing point of the transient and equilibrium reflectivity spectra. As $\gamma_\text{K}$ decreases, the crossing energy increases, evidencing an absence of an isosbestic point. Analogous maps acquired in the photo-doping regime [Fig.~2(h)] also exhibit a sign change. However, the crossing energy remains constant over an order of magnitude change in fluence, consistent with an isosbestic point. These measurements corroborate our simulations and highlight the unique distribution control afforded by Keldysh tuning. 
	
	\begin{figure}[t!]
		\centering
		\includegraphics[width=0.85\linewidth]{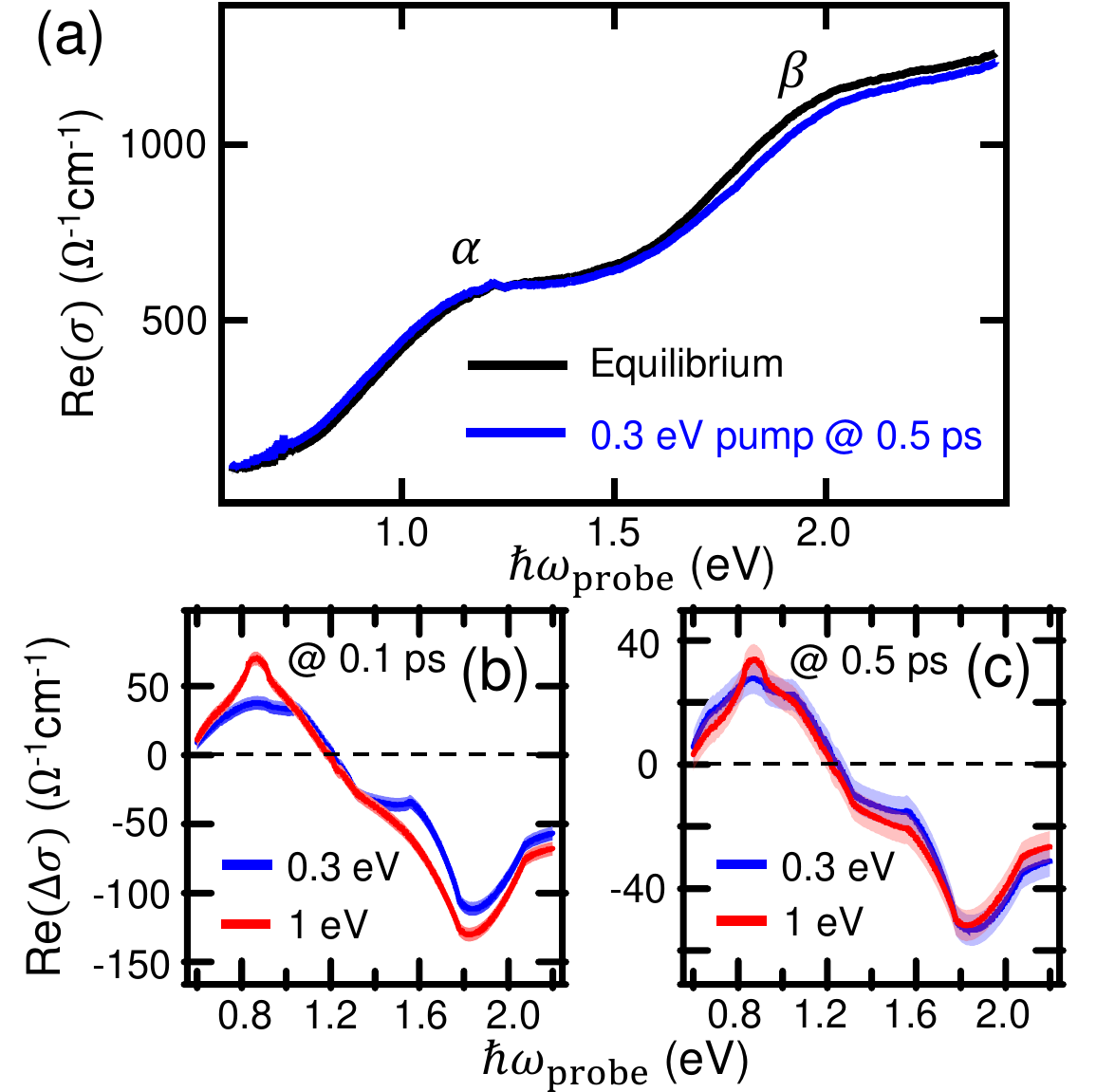}
		\caption{\small Non-equilibrium conductivity transients. (a)~Conductivity spectra of Ca$_2$RuO$_4$ in the un-pumped equilibrium state at 80 K and the 0.3 eV pumped non-equilibrium state at $t$ = 0.5 ps (fluence: 26~mJ/cm$^2$). (b)~Comparison of differential conductivity spectra between 0.3~eV pump ($\Delta\sigma_\text{0.3~eV}$) and scaled 1~eV pump ($A\Delta\sigma_\text{1~eV}$) cases at $t=0.1$~ps and (c) $t=0.5$~ps. Red and blue shades indicate error estimated from the $\omega_\text{probe}$-dependent fluctuations of the experimental $\Delta\sigma$ spectra.
		}
		\label{Fig3}
	\end{figure}
	
	To study the d-h thermalization dynamics in more detail, we used a Kramers-Kronig transformation to convert our differential reflectivity spectra into differential conductivity ($\Delta\sigma$) spectra \cite{SM}. Figure~3(a) shows the real part of the transient conductivity measured in the thermalized state ($t$ = 0.5 ps) following an 0.3 eV pump pulse of fluence $26$~mJ/cm$^2$ ($\gamma_\text{K}$ = 0.5), overlaid with the equilibrium conductivity. Subgap pumping induces a spectral weight transfer from the $\beta$ to $\alpha$ peak and a slight red-shift of the band edge, likely due to free carrier screening of the Coulomb interactions \cite{Golefmmodeheckzlsezi2015}. Unlike in the DC limit, there is no sign of Mott gap collapse despite $E_\text{pump}$ exceeding $10^9$ V/m. To verify that the electronic subsystem indeed thermalizes by $t=0.5$~ps, we compare the real parts of $\Delta\sigma_\text{0.3~eV}$ (fluence: $26$~mJ/cm$^2$) and $\Delta\sigma_\text{1~eV}$ (fluence: $4$~mJ/cm$^2$), the change in conductivity induced by subgap and above-gap pumping respectively, at both $t=0.1$ ps and 0.5~ps. A scaling factor $A$ is applied to $\Delta\sigma_\text{1~eV}$ to account for any differences in excitation density. As shown in Figure 3(b), the $t$ = 0.1 ps curves do not agree within any scale factor. This is expected because the linear and nonlinear pair production processes initially give rise to very different non-thermal distributions (Fig.~2). Conversely, by $t=0.5$~ps, the curves overlap very well [Fig.\,3(c)], indicating that the system has lost memory of how the d-h pairs were produced and is thus completely thermalized. 
	
	\begin{figure}[t!]
		\centering
		\includegraphics[width=0.85\linewidth]{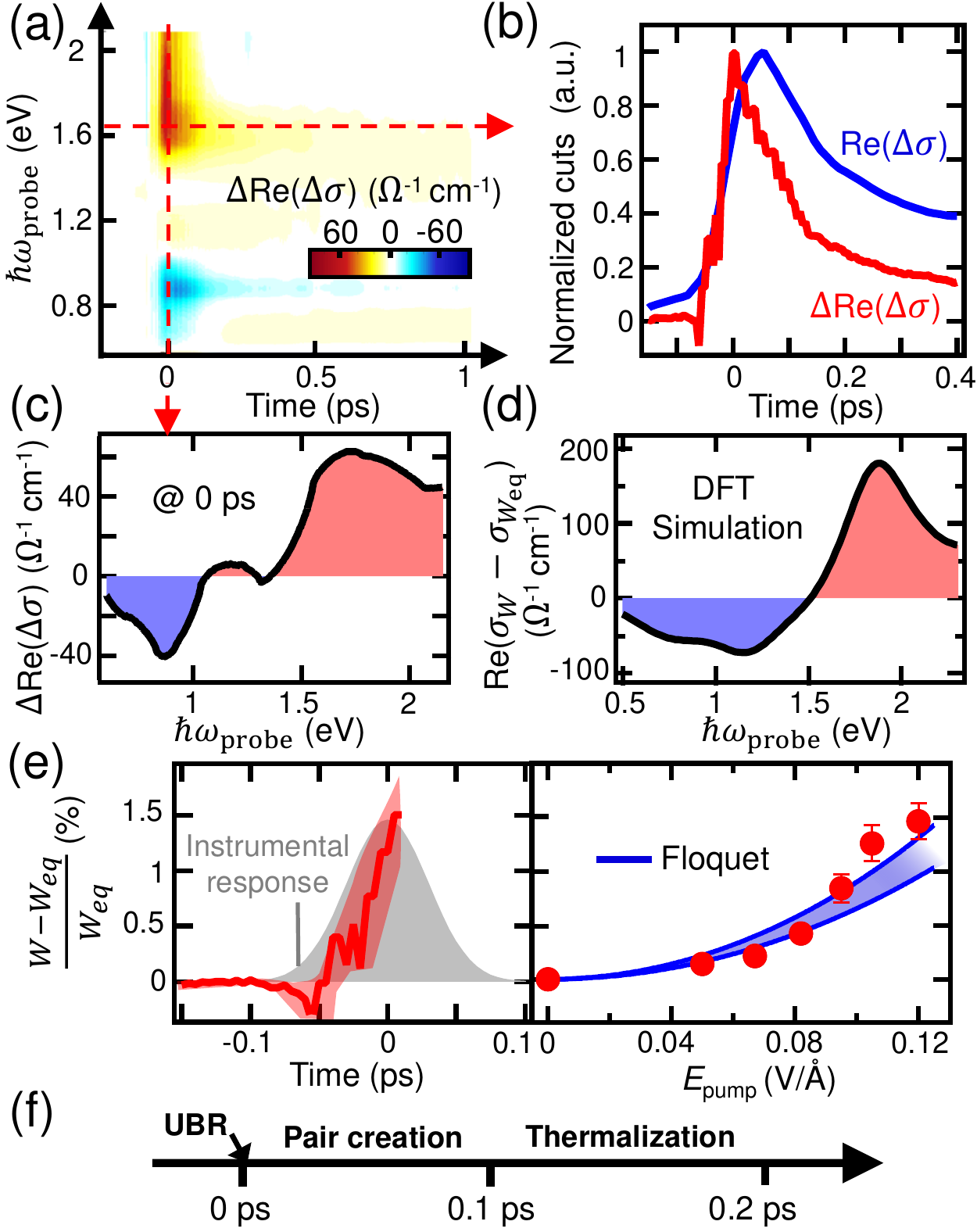}
		\caption{\small Ultrafast coherent bandwidth renormalization. (a)~$\Delta(\Delta\sigma)$ map obtained by subtracting scaled $\Delta\sigma_\text{1~eV}$ from $\Delta\sigma_\text{0.3~eV}$ spectra. (b)~A constant probe energy cut at 1.65~eV [dashed horizontal line in (a)] plotted together with $\Delta\sigma_\text{0.3~eV}$. (c)~A constant time cut at $t$ = 0 [dashed vertical line in (a)]. (d)~DFT simulation of the spectrum change induced by bandwidth broadening. $\sigma_W$ ($\sigma_{W_\text{eq}}$): conductivity with (without) bandwidth broadening. (e)~(left) Quantitative extraction of pump-induced bandwidth modification versus $t$ (with $E_\text{pump}=0.12$~V/\AA) and (right) versus $E_\text{pump}$ (with $t=0$~ps) based on fitting to DFT calculations. Red shaded region: error bar. Blue shaded region: Floquet theory prediction based on a periodically driven two-site cluster Hubbard model. Upper and lower bounds assume $U=3$~eV \cite{Jung2003} and $U=3.5$~eV \cite{Gorelov2010} respectively, where $U$ is the on-site Coulomb energy, with no other adjustable parameters. (f)~Chronology of non-thermal processes following an impulsive subgap drive.
		}
		\label{Fig4}
	\end{figure}

	Based on the observations in Figures\,3(b) and (c), the non-thermal window can be directly resolved by evaluating the time interval over which the quantity $\Delta(\Delta\sigma)=\Delta\sigma_\text{0.3~eV}-A\times\Delta\sigma_\text{1~eV}$ is non-zero \cite{SM}. Figure 4(a) shows the complete temporal mapping of $\Delta(\Delta\sigma)$ spectra. The signal is finite only around $t=0$~ps and is close to zero otherwise, supporting the validity our subtraction protocol. By taking a constant energy cut, we can extract a thermalization time constant of around 0.2 ps [Fig.~4(b)]. Interestingly, $\Delta R/R$ and $\Delta\sigma_\text{0.3~eV}$, which both track the d-h pair density, peak near 0.1 ps whereas $\Delta(\Delta\sigma)$ peaks earlier at $t$ = 0 when the d-h pair density is still quite low. This implies the existence of an additional coherent non-thermal process that scales with $E_\text{pump}$, which peaks at $t$ = 0, rather than with the d-h density.        
	
	To identify the physical process responsible for the $t$ = 0 signal, we examined how the electronic structure of Ca$_2$RuO$_4$ would need to change in order to produce the $\Delta(\Delta\sigma)$ profile observed at $t$ = 0 [Fig. 4(c)]. Using density functional theory (DFT), we performed an \textit{ab initio} calculation of the optical conductivity of Ca$_2$RuO$_4$ based on its reported lattice and magnetic structures below $T_\text{N}$. The tilt angle of the RuO$_6$ octahedra was then systematically varied in our calculation as a means to simulate a changing electronic bandwidth \cite{SM}. We find that both the real and imaginary parts of the measured $\Delta(\Delta\sigma)$ spectrum at $t$ = 0 are reasonably well reproduced by our calculations if we assume the bandwidth of the driven system ($W$) to exceed that in equilibrium $W_\text{eq}$ [Fig. 4(d)] \cite{SM}. This points to the coherent non-thermal process being a unidirectional ultrafast bandwidth renormalization (UBR) process that predominantly occurs under subgap pumping conditions [Fig. 4(f)].

	Coherent UBR can in principle occur via photo-assisted virtual hopping between lattice sites, which has recently been proposed as a pathway to dynamically engineer the electronic and magnetic properties of Mott insulators \cite{Mentink2015,Hejazi2019,Mikhaylovskiy2015,Batignani2015,Claassen2017,Wang2018}. To quantitatively extract the time- and $E_\text{pump}$-dependence of the fractional bandwidth change $(W-W_\text{eq})/W_\text{eq}$, we collected $\Delta(\Delta\sigma)$ spectra as a function of both time delay and pump fluence and fit them to DFT simulations \cite{SM}. As shown in Figure 4(e), the bandwidth change exhibits a pulse-width limited rise with a maximum $t$ = 0 value that increases monotonically with the peak pump field, reaching up to a relatively large amplitude of 1.5 \% at $E_\text{pump}$ = 0.12~V/\AA, comparable to the bandwidth increases induced by doping \cite{Fang2004} and pressure \cite{Nakamura2002}. Independently, we also calculated the field dependence of $(W-W_\text{eq})/W_\text{eq}$ expected from photo-assisted virtual hopping by solving a periodically driven two-site Hubbard model in the Floquet formalism \cite{Mentink2015,SM}, using the same model parameters for \CRO\ as in our Landau-Dykhne calculations [Fig. 1(a)]. We find a remarkable match to the data without any adjustable parameters [Fig. 4(e)]. Since bandwidth renormalization increases with the Floquet parameter $(eaE_\text{pump})/\hbar\omega_\text{pump}$ in the case of photo-assisted virtual hopping, where $a$ is the inter-site distance, this naturally explains why subgap pumping induces the much larger UBR effect compared to above-gap pumping.

	The ability to rationally tune a Mott insulator $in$ $situ$ over Keldysh space enables targeted searches for exotic out-of-equilibrium phenomena such as strong correlation assisted high harmonic generation \cite{Imai2020,Silva2018}, coherent dressing of quasiparticles \cite{Novelli2014}, Wannier-Stark localization \cite{Lee2014,Murakami2018}, AC dielectric breakdown \cite{Oka2012} and dynamical Franz-Keldysh effects \cite{Srivastava2004,TancogneDejean2020}, which are predicted to manifest in separate regions of Keldysh space. It also provides control over the nonlinear d-h pair production rate - the primary source of heating and decoherence under subgap pumping conditions - in Mott systems, which is crucial for experimentally realizing coherent Floquet engineering of strongly correlated electronic phases.

		We thank Swati Chaudhary, Nicolas Tancogne-Dejean, and Tae Won Noh  for useful discussions. The first-principles calculations in this work were performed using the QUANTUM ESPRESSO package. Time-resolved spectroscopic measurements were supported by the Institute for Quantum Information and Matter (IQIM), an NSF Physics Frontiers Center (PHY-1733907). D.H. also acknowledges support for instrumentation from the David and Lucile Packard Foundation and from ARO MURI Grant No.~W911NF-16-1-0361. X.L. acknowledges support from the Caltech Postdoctoral Prize Fellowship and the IQIM. G.C. acknowledges NSF support via grant DMR 1903888. M.C.L. acknowledges funding supports from the Research Center Program of IBS (Institute for Basic Science) in Korea (IBS-R009-D1). K.W.K. was supported by the National Research Foundation of Korea (NRF) grant funded by the Korea government (MSIT) (No. 2020R1A2C3013454). Y. M. was supported by the JSPS Core-to-Core Program No. JPJSCCA20170002 as well as the JSPS Kakenhi No. JP17H06136.

	
	%

	\newpage
	\onecolumngrid	

	\begin{center}
		\textbf{\large Supplemental Material for \\``Keldysh space control of charge dynamics in a strongly driven Mott insulator"}
	\end{center}


	\section{S1. Materials and methods}
	\subsection{A. Material growth and handling}
	
	Single crystals of \CRO\ used for time-resolved optical measurements were synthesized using a NEC optical floating zone furnace with control of growth environment \cite{Cao2000SM,Qi2012SM}. The lateral dimension of the crystal used in the optics measurement was roughly 0.5~mm by 1~mm. The crystal was freshly cleaved along the $c$-axis right before the measurements and immediately pumped down to pressures better than $7\times10^{-7}$ torr. \CRO\ is known to exhibit a violent metal-to-insulator transition accompanied by structural distortions at $T_\text{MIT}=357$~K. From \emph{in situ} optical microscopy measurements, we observed  that structural defects such as cracks frequently appear when temperature is swept across $T_\text{MIT}$. However, for the entire set of ultrafast optical experiments reported here, the measurement temperature was always kept in a range below $T_\text{MIT}$, for which crack formation can be avoided as long as the temperature ramping rate was slower than 1~K/minute.

	\subsection{B. Multicolor differential reflectivity measurements}
	
	We used a multicolor pump-probe setup to perform the differential reflectivity spectroscopy measurements. Our laser source was a Ti:sapphire-based regenerative amplifier (1~kHz, 5~mJ, 800~nm, 35~fs). The main output beam was split to feed two optical parametric amplifiers (OPAs) to produce pump and probe beams of different colors. The 0.3~eV MIR subgap pump beam was generated by a difference frequency generation (DFG) process in the non-collinear geometry between the signal and idler beams from the first OPA. The 1~eV above-gap pump beam was taken directly from the output from the same OPA after properly filtering out the idler component and the residual pump. The probe beam was generated from the second OPA equipped with a second-harmonic generation unit to cover the spectral range reported in Fig.\,1(c) of the main text. The pump was kept at normal incidence, so the electric field was in-plane. The pump and probe beams were always kept in cross-polarized geometry, and a linear polarizor was placed before the detector to filter out the pump scatter. Silicon and InGaAs detectors were used for the 1.4 - 2.2~eV and 0.5 - 1.4~eV probe photon energy ranges, respectively. All measurements were performed at 80~K unless mentioned otherwise. Note that for the 1~eV above-gap pump experiment, probe light with energies less than 1~eV will penetrate more into the sample than the pump. This type of pump-probe penetration depth mismatch has been accounted for when calculating $\Delta\sigma_\text{1~eV}$ through the KK transform, and is found to have little impact on the entire analysis.

	\subsection{C. Density functional theory simulations}
	
	The electronic structure of \CRO\ was calculated from first principles with density functional theory implemented in the QUANTUM ESPRESSO package. The calculation used a plane-wave basis set and scalar relativistic  norm-conserving Vanderbilt pseudopotentials. The energy cutoff was set to 60~Ry. A self-consistent calculation using a 4$\times$4$\times$4 Monkhorst-Pack grid was run at first, followed by a non-self-consistent calculation with a denser user defined grid. Convergence was tested for the energy cutoff and the grid density. The calculation was set to the spin-polarized mode to take into account the low-temperature antiferromagnetic structure of \CRO, and to the DFT+$U$ mode to take into account the Coulomb correlation. The real and imaginary parts of the optical conductivity were calculated by the epsilon.x package after the non-self-consistent calculation. Finite interband and intraband smearings were used to avoid sharp spikes in the spectra caused by numerical issues.
	
	To unambiguously confirm the modification of the electronic structure made by the pump field, we performed calculations using different input material parameters of \CRO\ to account for different scenarios. The UBR scenario was implemented by changing the crystal structural input by tuning the tilting angles of the RuO$_6$ octahedra; the tilting angle changes the Ru-O-Ru bond angle, and thereby modifies the bandwidth. The case of Hubbard-$U$ modification was simulated by directly tuning the $U$ value in the DFT+$U$ input. Details of the simulation results and how we chose different input parameters of \CRO\ to simulate different cases are discussed in detail in the Section S6.

	\section{S2. Differential reflectivity spectra}
	
	\subsection{A. Fitting analysis}
	
\begin{figure}[h]
		\centering
		\includegraphics[width=0.7\linewidth]{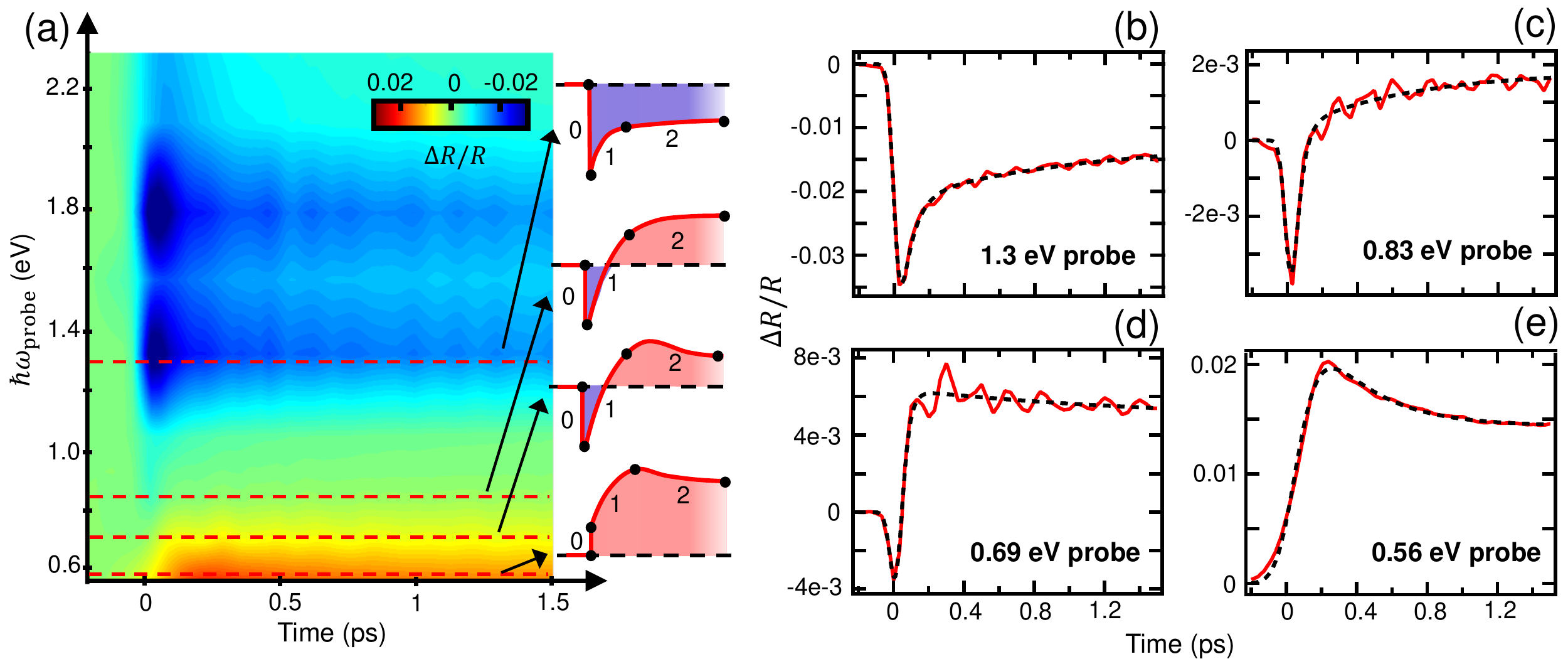}
		\caption{(a)~$[\Delta R/R](t)$ map for 0.3~eV pump and $F=$15.2~mJ/cm$^{2}$. Transients of four types are schematically shown to the right. (b)-(e)~Four constant probe photon energy cuts marked by red dashed lines in (a). Red lines are data. Black dashed lines are fittings from the convolution.}
		\label{dRRfitting}
	\end{figure}

	Here we discuss the analysis procedure for the differential reflectivity data reported in this paper. $[\Delta R/R](t)$ transients at ten consecutive probe photon energies were measured within the range of 0.55 - 2.2~eV, forming the three-dimensional colormaps in Fig.\,2(e), (f) in the main text. Figure\,\ref{dRRfitting}(a) shows another example of a colormap of similar type for 0.3~eV pump at a fluence ($F$) of 15.2~mJ/cm$^{2}$.  
	
	For all probe energies, the dynamics of $[\Delta R/R](t)$ transients can be described by three consecutive steps as expressed in the double exponential function
	\begin{equation}
	f(t)=\begin{cases}
	0, & \text{for}\ t<0\ \text{ps}\\
	A_{1}e^{-t/t_{1}}+A_{2}e^{-t/t_{2}}+C, & \text{for}\ t\geq0\ \text{ps},
	\end{cases}
	\label{dblexpdcy}
	\end{equation}
	namely, the initial excitation upon pump arrival at $t=0$~ps (Step 0), followed by a fast exponential process with $t_1\sim0.1$~ps (Step 1), and a slow exponential process with $t_2\sim1$~ps (Step 2) which settles the signal down to a constant plateau that decays on much longer time scales. However, depending on the probe photon energy, the signs and magnitudes of $A_1$, $A_2$, and $C$ can change, leading to four types of traces as schematically summarized in  Fig.\,\ref{dRRfitting}(a). The signs of the coefficients for different probe energies are summarized in Table\,\ref{parameters}. Four horizontal cuts to the experimental data marked by red dashed lines in Fig.\,\ref{dRRfitting}(a) are shown in Fig.\,\ref{dRRfitting}(b)-(e) to represent these four types of transients.
	
	\begin{table}[htb]
		\begin{tabular}{|c|c|c|c|c|}
			\hline
			& $A_1+A_2+C$ & $A_1$ & $A_2$ & $C$ \\\hline
			1.2 - 2.2 eV &    $<0$     &  $<0$  &  $<0$  & $<0$  \\\hline
			0.8 - 1.1 eV &     $<0$    &  $<0$  &  $<0$  & $>0$  \\\hline
			0.7 eV       &    $<0$     &  $<0$  &  $>0$  &  $>0$ \\\hline
			0.55 - 0.7 eV       &     $>0$    &  $<0$  &  $>0$  &  $>0$ \\
			\hline
		\end{tabular}
		\caption{Signs of coefficients for different probe energies.}
		\label{parameters}
	\end{table}
	
	Then, we fit the experimental $[\Delta R/R](t)$ with a convolution $[f\otimes IRF](t)$, where $f(t)$ is the intrinsic dynamics (Eq.\,\ref{dblexpdcy}), and the instrumental response function (IRF) takes the form of a Gaussian
	\begin{equation}
	IRF(t)=\frac{1}{\sigma\sqrt{2\pi}}\text{exp}[-\frac{-(t-t_0)^2}{2\sigma^2}],
	\label{IRF}
	\end{equation}
	where $\sigma$ is the instrumental time resolution and $t_0$ is the time zero of the measurement at which the pump and probe pulses reach temporal overlap. Fitting parameters are $A_1$, $t_1$, $A_2$, $t_2$, $C$, $t_0$, and $\sigma$. 
	
\begin{figure}[h]
		\centering
		\includegraphics[width=0.5\linewidth]{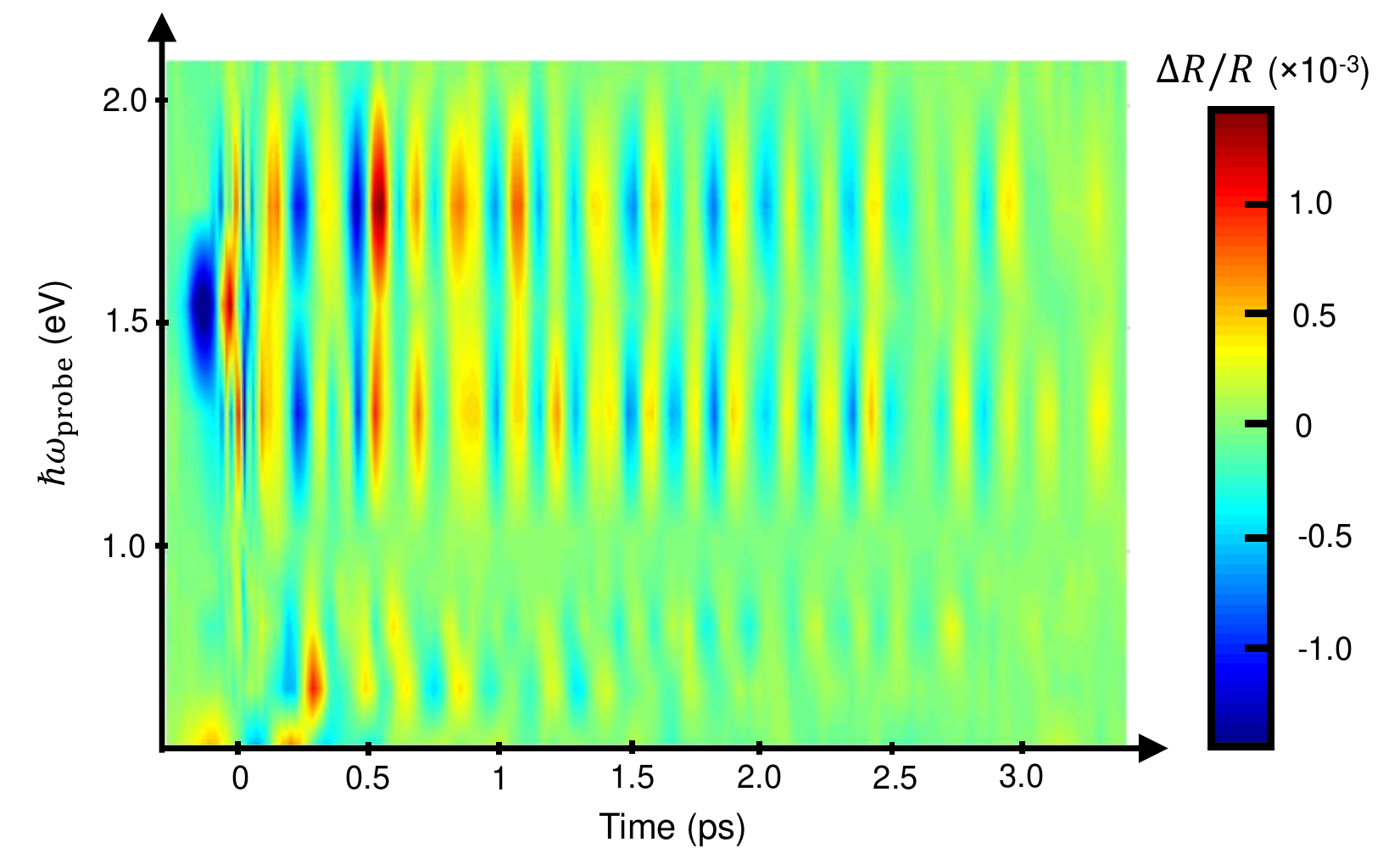}
		\caption{Coherent phonon oscillation map obtained by subtracting the $[\Delta R/R](t)$ map in Fig.\,\ref{dRRfitting}(a) with the fitted electronic background.}
		\label{Phononalign}
	\end{figure}

	The black dashed lines in Fig.\,\ref{dRRfitting}(b)-(e) are the fitted curves, which are in close agreement with the experimental transients. There is a slight change in $\sigma$ depending on pump and probe photon energies, ranging from $50$~fs to $120$~fs. The fitted $\sigma$ values are used to infer the pump pulse widths. The pulse duration for 0.3~eV pump is around $\Delta t=100$~fs, which is used for estimating the pump electric field strength $E_\text{pump}$ through the expression for the energy density $u=\frac{1}{2}\varepsilon_0 (E_\text{pump})^2=F/(c\Delta t)$, where $F$ is the fluence in vacuum and $c$ is the speed of light. Determination of $t_0$ allows us to align transients at different probe energies temporally with a common time zero. Robustness of the fitted $t_0$ can be seen in the coherent phonon oscillation map in Fig.\,\ref{Phononalign}. The map is obtained by subtracting the $[\Delta R/R](t)$ map (temporally aligned with $t_0$) with the fitted electronic backgrounds. Good alignment of the oscillation phase of the coherent phonon across the entire probe energy range suggests that the fitting procedure for $t_0$ is reliable.
	
	\subsection{B. Time dynamics of zero-crossing feature}
	
	In Fig.\,2 of the main text, the evolution of the zero-crossing feature of the $[\Delta R/R](t)$ spectra versus pump fluence is used as a metric to distinguish the subgap Keldysh tuning scenario and the above-gap photodoping scenario. What is visually unclear when showing colormaps as in Figs.\,2(e)-(h) is the evolution of the energy of the zero crossing features after $t=0.1$~ps, the time delay we identified to have the highest d-h pair density and most nonthermal distribution. Here, we show the time dynamics of the zero-crossing feature a lot more clearly by plotting $|[\Delta R/R]|(t)$, the absolute value of $[\Delta R/R](t)$ on a logarithmic scale.
	
\begin{figure}[h]
		\centering
		\includegraphics[width=0.7\linewidth]{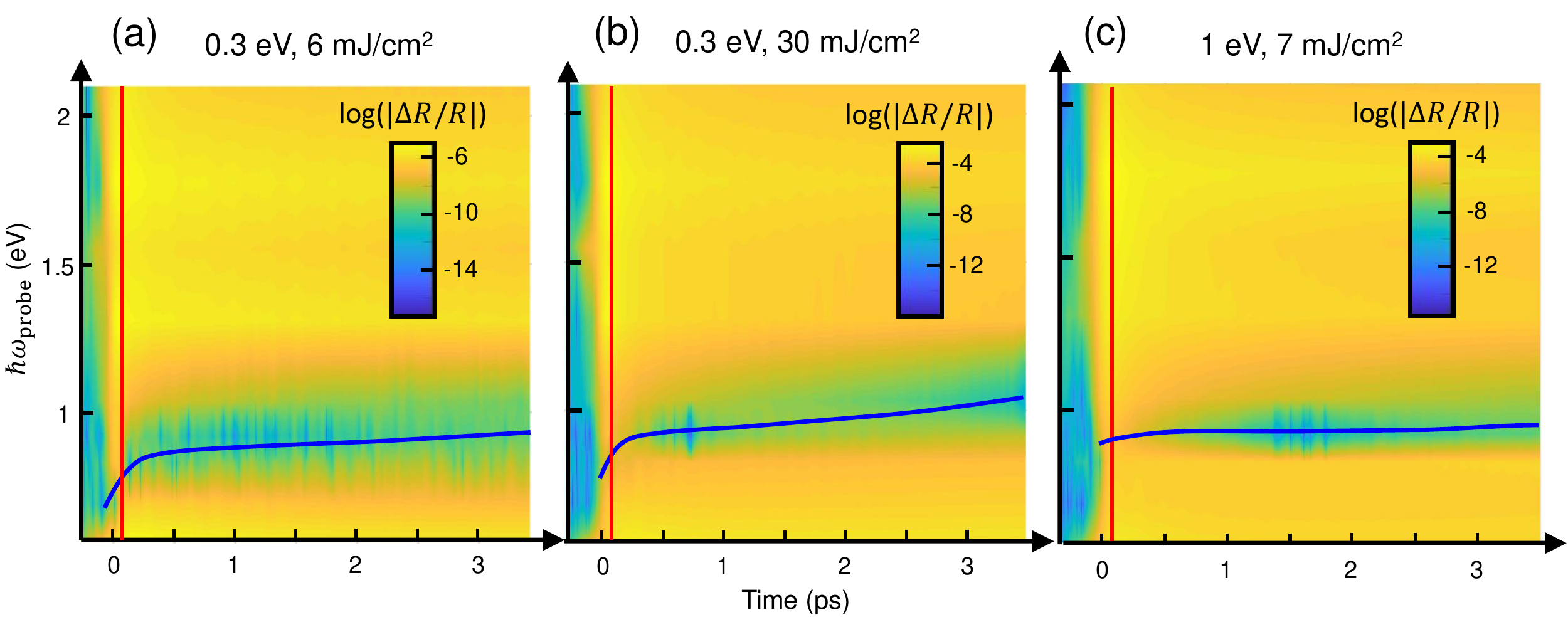}
		\caption{Time dynamics of the zero-crossing feature highlighted in the logarithmic plot of $|[\Delta R/R]|(t)$ map. (a) and (b),~Two fluences for the 0.3~eV pump scenario. (c)~Highest fluence for the 1~eV pump scenario. Blue lines are guides to the eye for the minimum of $|[\Delta R/R]|(t)$, highlighting the shift of zero-crossing versus time. Red lines mark the cuts at $t=0.1$~ps, where the pair distribution is nonthermal.}
		\label{logzerocrossing}
	\end{figure}

	Figure~\ref{logzerocrossing} shows three representative maps of $|[\Delta R/R]|(t)$ with different pumping conditions. The zero crossing feature is where $|[\Delta R/R]|(t)$ is smallest, corresponding to the valleys marked by blue lines in the graphs. The feature clearly continues to shift in energy after $t=0.1$~ps (red vertical cuts), indicating that the optical response of the sample undergoes subsequent stages of evolution, including pair thermalization, interband recombination, and heating, each with a characteristic timescale. A notable example is at $t=3$~ps, where pairs have mostly recombined and the electronic and the lattice systems have equilibrated at a higher transient temperature; the energy of the zero-crossing is an indicator of sample heating \cite{Lee2018SM}. The fluence of Fig.~\ref{logzerocrossing}(b) is higher, and creates more heat than that in Fig.~\ref{logzerocrossing}(a), which naturally explains why its zero-crossing is at higher energy at $t=3$~ps.

	\section{S3. Calculation of the Keldysh parameter space}
	
	The Keldysh parameter space is formed by the subgap pump field strength and pump photon energy. In Figs.~1 and 2 of the main text, we plot the total d-h pair production rate $\Gamma$ in the Keldysh space as well as the momentum dependent transition probability $P_p$. These calculations were obtained by using the Landau-Dykhne method combined with the Bethe Ansatz, as reported in Ref.\,\cite{Oka2012SM}.

	\subsection{A. The Landau-Dykhne method}
	
	The Landau-Dykhne method combined with the Bethe Ansatz has been used to model the nonlinear d-h pair production process in Mott insulators across the entire Keldysh parameter space, from the multiphoton regime to the tunneling regime. We closely followed the procedure developed by Oka \cite{Oka2012SM}; the theory was applied to a 1-dimensional (1D) Hubbard model in the original paper, but results and equations therein have been widely referenced by dielectric breakdown experiments on materials with higher dimensions \cite{Mayer2015SM,Yamakawa2017SM}. Therefore, we anticipate that the model can provide important qualitative guidance to our experiments, even though the 1D Hubbard model cannot fully reflect the realistic electronic structure or multiband Mott nature of \CRO.
	
	For a 1D Hubbard model in a time-dependent electric field, the adiabatic perturbative theory expands the time-dependent state into the linear combination of adiabatic eigenstates
	\begin{equation}
	\ket{\Psi(t)}=a(t)\ket{0;\Phi(t)}+b(t)\ket{p;\Phi(t)}_{dh},
	\end{equation}
	where $p$ is momentum, $\Phi(t)$ is the Peierls phase, and $a(t)$ and $b(t)$ are the probability amplitudes for the channel at $p$ to be in the ground state (no pair) or in the excited state (with a pair). The $p$-dependent transition probability $P_p = [b(t)]^2$ can be calculated as 
	\begin{equation}
	P_p=\text{exp}(-2\text{Im}\mathcal{D}_p),
	\end{equation}
	where $\mathcal{D}_p$ is the difference between the dynamical phase of the ground state and the excited state. After more treatments, Ref.\,\cite{Oka2012SM} gives
	\begin{equation}
	\text{Im}\mathcal{D}_p = \text{Im}\mathcal{D}_{p1}+\text{Im}\mathcal{D}_{p2},
	\end{equation}
	and
	\begin{align}
		\text{Im}\mathcal{D}_{p1} & =\int_{p}^{0}\Delta E(l)\text{Im}\left(\frac{1}{F(p-l)}\right) dl,\\
		\text{Im}\mathcal{D}_{p2} & =\int_{0}^{1/\xi}\Delta E(il)\text{Im}\left(\frac{i}{F(p-il)}\right) dl.
	\end{align}
	Here, $\Delta E$ is the gap function, $F(\Phi)=\pm\sqrt{F_0^2-\Omega^2\Phi^2}$ is the time-dependent field with sinusoidal oscillations, $\xi$ is the d-h correlation length, $\Omega$ is the pump frequency, and $F_0$ is the amplitude of $F$. After $P_p$ is calculated, the total d-h pair production rate can be obtained by an integral
	\begin{equation}
	\Gamma=\frac{\Omega}{2\pi}\int_{-\pi}^{\pi}\frac{dp}{2\pi}P_p.
	\end{equation}
	
	\begin{figure}[h]
		\centering
		\includegraphics[width=0.8\linewidth]{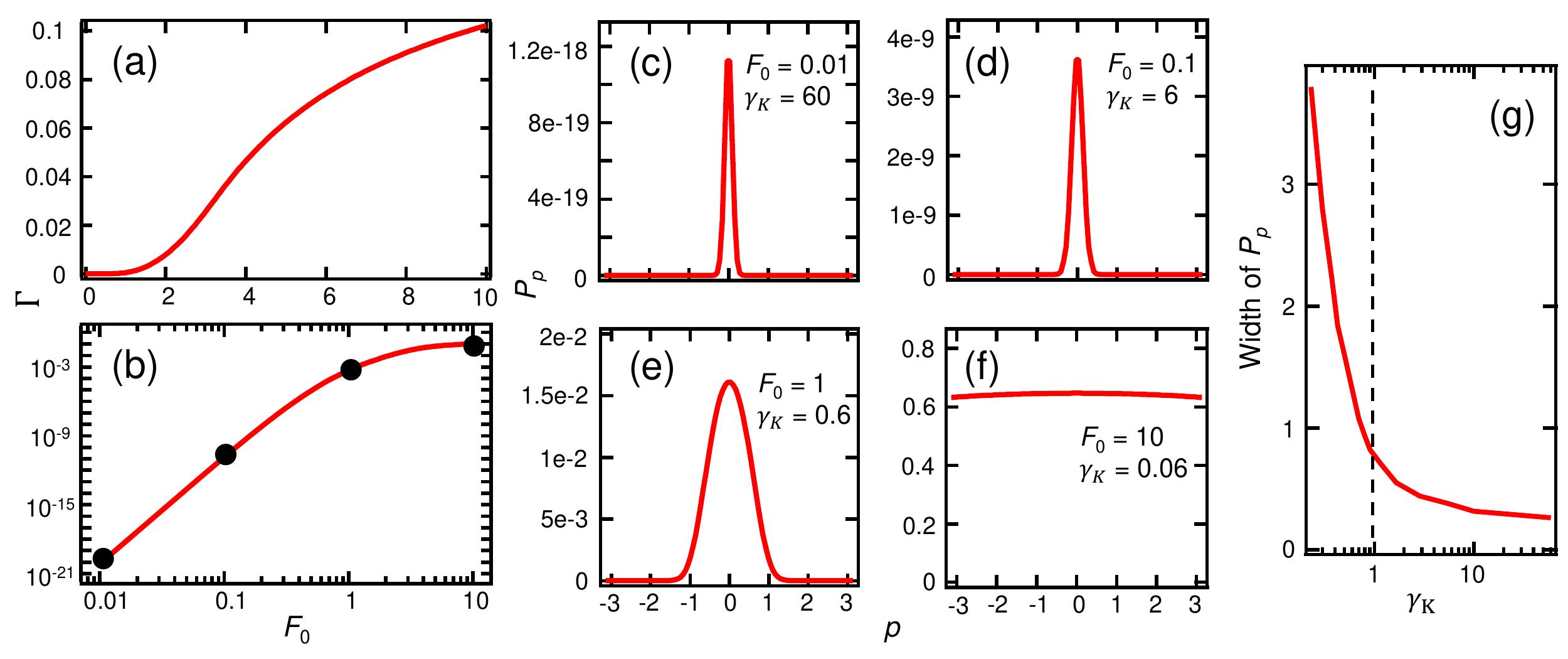}
		\caption{Field-dependent nonlinear excitation for $U=8$, $\Delta = 4.6$, $\Omega = 1$. (a)~Total d-h production rate $\Gamma$ versus field strength $F_0$. (b)~Same as (a) but on a logarithmic scale. (c)-(f)~Momentum dependent transition probability $P_p$ at the four different field strengths marked with black circles in (b). The corresponding Keldysh parameters are also labeled. (g)~Width of $P_p$ versus $\gamma_K$.}
		\label{Okarepro}
	\end{figure}
	
	Figure~\ref{Okarepro} shows a validating calculation using  Hubbard $U=8$, Mott gap $\Delta = 4.6$, pump frequency $\Omega = 1$, at various field strengths $F_0$. The energy unit is the hopping energy $t_\text{hop}$, $F_0$ is in the unit of $t_\text{hop}/a$, where $a$ is the lattice parameter. The nonlinear production rate $\Gamma$ versus $F_0$ is shown in Figs.\,\ref{Okarepro}(a) and (b) on linear and logarithmic scales, respectively; in (b), the Keldysh crossover is observed as the line deviates from the power law scaling of the multiphoton process as $F_0$ increases. $P_p$'s at the four representative $F_0$'s marked by black circles in Fig.\,\ref{Okarepro}(b) are shown in Figs.\,\ref{Okarepro}(c)-(f). Drastic broadening of $P_p$ is clearly seen in the vicinity of the Keldysh crossover $\gamma_K\sim 1$, while in the deep multiphoton regime ($\gamma_K\gg 1$) the broadening effect is minimal; see Fig.\,\ref{Okarepro}(g), and the comparison between Figs.\,\ref{Okarepro}(c) and (d). These results all well reproduce findings in Ref.\,\cite{Oka2012SM}.
	
	\subsection{B. Band parameters of \CRO\ used in the calculation}
	
	Band parameters of \CRO\ and realistic subgap pumping conditions were plugged in the equations above to calculate the Keldysh map and the $P_p$ curves in Figs.\,1 and 2 of the main text. We used the parameters reported in the dynamical mean-field theory calculations for \CRO\ \cite{Gorelov2010SM}, with $U=3.5$~eV, $\Delta = 0.6$~eV (from optical measurements in \cite{Jung2003SM}), $a=5.6$ \AA\ (in-plane lattice parameter), and $t_\text{hop} = 0.23$~eV (the hopping integral between $xy$ orbitals, since joint density of states near the Mott gap is mostly contributed by $d_{xy}\to d_{xz/yz}$ transitions). The correlation length \cite{Stafford1993SM} can be calculated with $\xi=[\ln(U/4.377t_\text{hop})]^{-1}a$ in the strong-coupling limit (which holds for \CRO\ since $U/t_\text{hop}=15$ \cite{Stafford1993SM}). We estimate $\xi=4.45$~\AA. This value is the same order of magnitude as $\xi=2.1$~\AA\ estimated for VO$_2$ \cite{Mayer2015SM}, which is another Mott insulator showing a cooperative charge-lattice response across a temperature-driven metal-to-insulator transition.

	\section{S4. Simulation of optical properties of a photoexcited insulator}
	
\begin{figure}[h]
		\centering
		\includegraphics[width=0.8\linewidth]{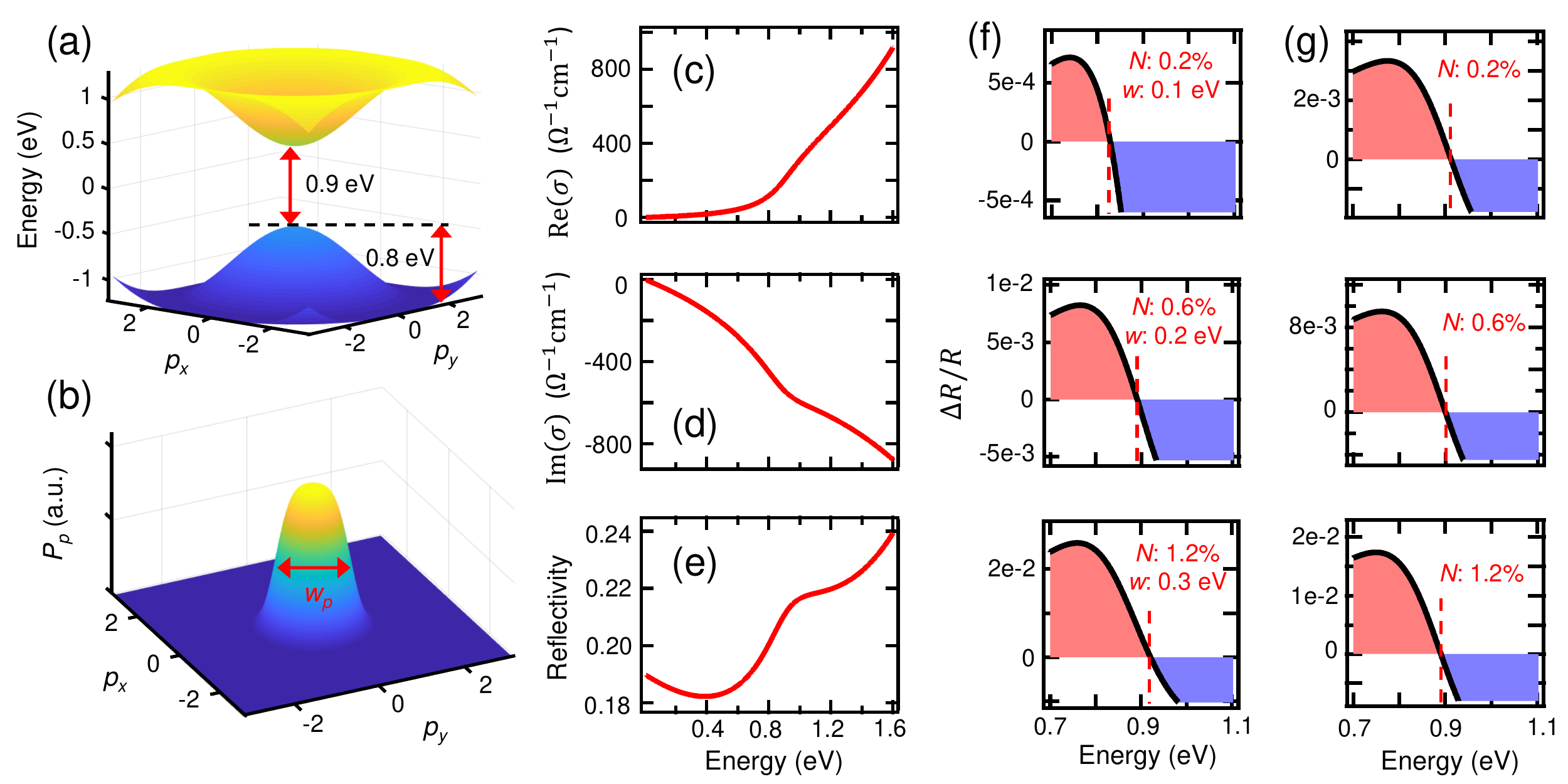}
		\caption{Simulation of optical properties of a photoexcited insulator. (a)~Band structure. (b)~Momentum dependent photocarrier distribution. $w_p$ ($w$) represents the width of the distribution in the momentum (energy) space. (c)-(e)~Equilibrium optical properties calculated from the band structure in (a) with no photocarriers. (f)~Differential reflectivity spectrum (spectrum with photocarriers subtracted by that without photocarriers) at various carrier densities ($N$) and width of distribution $w$. (g)~Same as (f) except that the $P_p$ is adjusted to the 1-eV-pump nonthermal distribution. $N$ changes consistently, while the width of distribution stays constant for the three panels.}
		\label{bandsim}
	\end{figure}

	To understand how photo-induced band filling affects the $\Delta R/R$ spectrum, we performed simulations assuming a simplified insulator model. Figure~\ref{bandsim}(a) shows the band structure of the simulated insulator. Due to the quasi-2D nature of the electronic structure of \CRO, we considered a cosine-type dispersion in the 2D momentum plane formed by $p_x$ and $p_y$. Conduction and valence bands are symmetric about zero energy, each with a bandwidth of 0.8~eV, and separated by a direct gap of 0.9 eV (which, upon considering the band edge smearing effect due to dephasing, gives an optical gap of 0.6~eV). These parameters were chosen to produce similar band-edge optical properties as \CRO. The optical susceptibility spectrum resulting from interband transitions can be obtained by \cite{Rosencher2002SM}
	\begin{equation}
	\chi(\omega)=\sum_{p_x, p_y}\frac{e^2x_{vc}(p_x, p_y)T_2}{\epsilon_0\hbar}\frac{[\omega_\text{probe}-\Delta(p_x, p_y)/\hbar]T_2-i}{[\omega_\text{probe}-\Delta(p_x, p_y)/\hbar]^2T^2_2+1}[N_v(p_x, p_y)-N_c(p_x, p_y)],
	\label{optsusc}
	\end{equation}
	where $e$ is the electron charge, $\epsilon_0$ is vacuum permittivity, $\hbar$ is Planck's constant divided by $2\pi$, $x_{vc}$ is the matrix element of the vertical interband transition at a particular momentum (assumed to be a constant for all momenta for simplification), $T_2$ (assume to be constant) is the band dephasing time, $\omega_\text{probe}$ represents probe frequency, $\Delta(p_x, p_y)$ represents the gap energy, and $N_v$ and $N_c$ are the electron occupations of the valence and conduction bands, respectively. The physical picture of the equation is that the bands are viewed as an ensemble of vertical two-level systems (TLSs) in the $p_x$- $p_y$ plane with level separations $\Delta(p_x, p_y)=\hbar(\omega_{c}-\omega_{v})$; each TLS contributes a Lorentzian oscillator, weighted by its corresponding occupation factor, to the total susceptibility.
	
	In equilibrium ($N_v=1$, $N_c=0$ for all $p_x$ and $p_y$), we calculated the susceptibility by Eq.\,\ref{optsusc} and converted it into static real and imaginary parts of conductivity $\sigma$ and reflectivity, as shown in Figs.\,\ref{bandsim}(c)-(e). The values and trends of the curves are similar to those of \CRO\ measured around its $\alpha$ peak onset energy (Mott band edge), but the higher energy transitions that involve multiple orbitals in \CRO, such as the $\beta$ and $\gamma$ peaks, are not accounted for by the model.
	
\begin{figure}[h]
		\centering
		\includegraphics[width=0.3\linewidth]{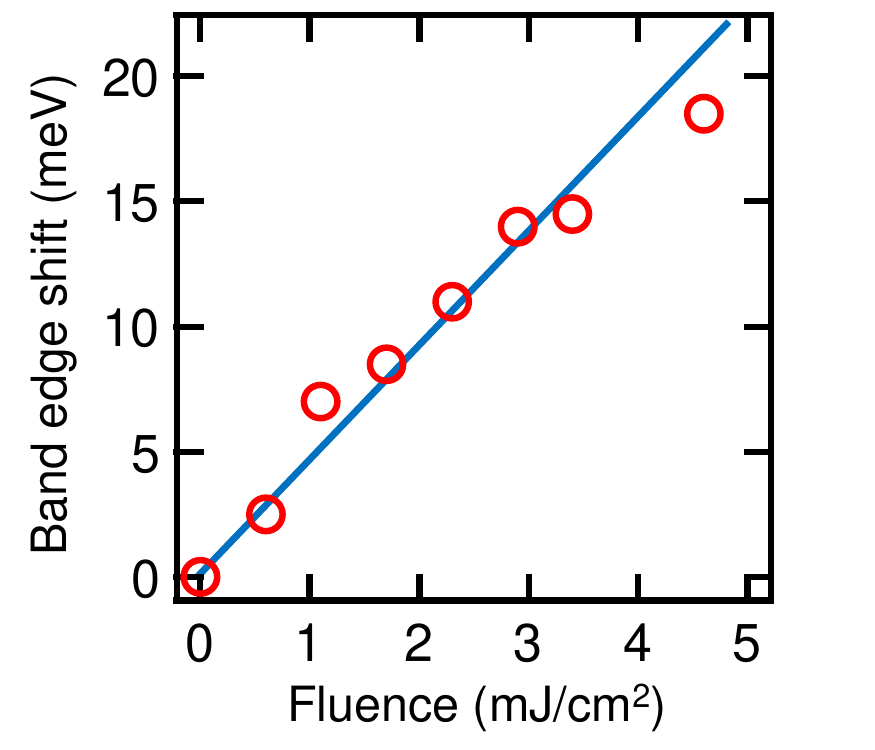}
		\caption{Band edge energy redshift versus fluence for 1~eV pump experiments. This graph is obtained by performing a Kramers-Kronig transform detailed in the next section.}
		\label{Gapshift}
	\end{figure}

	Next, we consider the laser-driven case. We used the Gaussian distribution to account for a total of $N$ photoexcited nonthermal carriers
	\begin{equation}
	N=\sum_{p_x, p_y}N_c(p_x, p_y)=\sum_{p_x, p_y}\frac{A}{w\sqrt{2\pi}}\exp{\frac{-[\omega_{c}(p_x, p_y)-\omega_0]^2}{2w^2}},
	\label{ocptGaussian}
	\end{equation}
	where we specified width $w$, peak center $\omega_0$, and $N$ to determine $A$ and $N_c(p_x, p_y)$; for 0.3~eV pump experiments, we set $\omega_0$ to be half of the direct gap (assuming zero energy centers the gap), and $w$ progressively increases with $N$ to mimic the width evolution of the $\gamma_K$-dependent $P_p$ distribution obtained from the Landau-Dykhne theory, while for 1~eV pumping, we set $\omega_0=0.5$~eV, and $w$ remains constant with increasing $N$. A representative Gaussian distribution landscape is shown in Fig.\,\ref{bandsim}(b), mimicking the situation for subgap pumping at a relatively low fluence, where the states at the band edge (where the gap is smallest) are mostly occupied. The nonthermal photocarrier distribution affects $N_c(p_x, p_y)$ and $N_v(p_x, p_y)=1-N_c(p_x, p_y)$, and therefore, modifies the nonequilibrium $\sigma$ and reflectivity. By applying the Fresnel equations, we simulated $\Delta R/R$ spectra for various photocarrier densities $N$ and distribution widths $w$, and plotted three scenarios in Fig.\,\ref{bandsim}(f); $N$ is expressed as the percentage of the pair density within the maximum allowed number of pairs in the bands. One detail we noticed was that simply considering the filling-induced optical bleaching will only lead to negative $\Delta R/R$ for all probe energies. This is because the equilibrium reflectivity develops a peak structure around the energy where the gap onsets (see Fig.\,\ref{bandsim}(c) and (e)), and filling will only bleach the gap feature, and therefore, lead to a suppression of the reflectivity peak. We found that, to match the experimental fact that positive $\Delta R/R$ regions are present in the experimental data, a term considering the photocarrier-induced band edge redshift has to be included. Therefore, we assumed the photoinduced change in $\Delta$ to be proportional to $N$ (If the change is small, only the linear term in the Taylor expansion is retained.). This assumption is reasonable because quantitative extraction of band edge energy shift versus fluence in the 1~eV pump experiments indeed recovers linearity; see Fig.\,\ref{Gapshift}. 
	
	For the three scenarios plotted in Fig.\,\ref{bandsim}(f), we included both the filling effect through $N_v(p_x, p_y)$ and $N_c(p_x, p_y)$ and the band edge redshift through $\Delta$. For the three rows in Fig.\,\ref{bandsim}(f), the simultaneous increase of $N$ and $w$ is for simulating the fluence dependence of our subgap pumping experiment, where pair density increases with fluence, and $w$ increases as $\gamma_K$ decreases (predicted by the Landau-Dykhne theory). An apparent expansion of the $\Delta R/R>0$ region is observed from the top panel to the bottom panel. This key feature is present in the simulations and experimental data in Fig.\,2 of the main text, because this type of evolution of $\Delta R/R$ prohibits an isosbestic point in the nonequilibrium reflectivity spectra in the subgap pumping scenario. On the other hand, the simulations shown in Fig.\,\ref{bandsim}(g), which are adapted to the experimental 1 eV pumping condition, and accounts for both the increase of $N$ and the band edge redshift ($\propto N$) but not the increase of $w$ or any change in the nonthermal probability distribution function, fails to reproduce the expansion of the $\Delta R/R>0$ region. The fact that the entire $\Delta R/R$ spectrum seems to scale with $N$ for a constant distribution function in Fig.\,\ref{bandsim}(g) leads to the appearance of an isosbestic point in the nonequilibrium reflectivity spectra for the 1 eV pumping case; which was exactly observed in experiments.

	\section{S5. Kramers-Kronig transform and differential optical conductivity}
	
	Figures~3 and 4 of the main text shows differential optical conductivity spectra ($\Delta\sigma$), which were obtained by a Kramers-Kronig (KK) analysis of the $\Delta R/R$ spectra. In this section, we first discuss our KK algorithm, which converts experimental $\Delta R/R$ spectra within a limited probe energy range to $\Delta\sigma$, provided that the static broadband optical conductivity $\sigma$ is known. Second, we will discuss details of identifying modifications to differential conductivity for 0.3~eV pumping ($\Delta\sigma_\text{0.3~eV}$) by Floquet bandwidth renormalization.
	
	\subsection{A. The regional KK transform algorithm}
	
	KK transform is a powerful technique that enables calculation of intrinsic complex-valued optical constants of a material from the reflectivity data alone. If the reflectivity spectrum $R(\omega)$ is known, the reflection phase $\theta(\omega)$ can be calculated as
	\begin{equation}
	\theta(\omega)=\frac{1}{\pi}\int_{0}^{\infty}\text{ln}\abs{\frac{\omega^{'}+\omega}{\omega^{'}-\omega}}\frac{d\text{ln}\sqrt{R(\omega^{'})}}{d\omega^{'}}d\omega^{'}
	\label{theta}
	\end{equation}
	without directly measuring it in experiments, and the real and imaginary parts of refractive index can be calculated by
	\begin{align}
		\text{Re}(n) & =\frac{1-R}{1+R-2~\text{cos}\theta\sqrt{R}}\\
		\text{Im}(n) & =\frac{-2~\text{sin}\theta\sqrt{R}}{1+R-2~\text{cos}\theta\sqrt{R}}.
	\end{align}
	
	And the the optical conductivity $\sigma$ can be obtained by $\sigma=(n^2-1)\omega\epsilon_0/i$. However, to use Eq.\,\ref{theta}, $R(\omega)$ must be known from zero to infinite frequencies, which is impractical for experiments. Various methods exist that extrapolate $R(\omega)$ within a limited measurement range to high and low frequencies to complete the calculation.
	
	Our situation is the following. The static optical constants of \CRO\ without optical pumping have already been determined by measuring broadband $R(\omega)$ from 80~meV to 6.5~eV, followed by data extrapolation and KK transform. However, the key issue is that our pump-probe measurement that gives $\Delta R/R$ covers a smaller frequency range (0.5~eV to 2.2~eV), and we hope to obtain $\Delta\sigma$ in the same range. Equation\,\ref{theta} cannot be directly applied because no model exists to extrapolate $\Delta R/R$. But Eq.\,\ref{theta} shows a strong resonance at $\omega\sim\omega^{'}$, suggesting that frequencies that are away from the range of interest contribute less to $\theta$. And when $\Delta R/R\ll1$, it is possible that, numerically, simply considering the $\Delta R/R$ only in the measurement range is accurate enough to give $\Delta\sigma$ in the same range. We followed the discussions in Ref.\,\cite{Roessler1965SM} to perform such a regional KK analysis.
	
	The $\theta$ integral can be written as the sum of three frequency ranges, namely, the low-frequency range, the measurement range, and the high-frequency range:
	\begin{equation}
	\theta(\omega)=-\frac{1}{\pi}\int_{0}^{0.5~\text{eV}}f(R,\omega)d\omega^{'}-\frac{1}{\pi}\int_{0.5~\text{eV}}^{2.2~\text{eV}}f(R,\omega)d\omega^{'}-\frac{1}{\pi}\int_{2.2~\text{eV}}^{\infty}f(R,\omega)d\omega^{'},
	\label{threeterms}
	\end{equation}
	where
	\begin{equation}
	f(R,\omega)=\text{ln}\sqrt{R(\omega^{'})}\frac{d}{d\omega^{'}}\left(\text{ln}\abs{\frac{\omega^{'}+\omega}{\omega^{'}-\omega}}\right).
	\end{equation}
	Applying the generalized mean value theorem to first and third integrals in Eq.\,\ref{threeterms}, and defining the second term as $\phi$ gives
	\begin{equation}
	\theta(\omega)=A~\text{ln}\abs{\frac{0.5~\text{eV}+\omega}{0.5~\text{eV}-\omega}} + \phi(\omega) + B~\text{ln}\abs{\frac{2.2~\text{eV}+\omega}{2.2~\text{eV}-\omega}},
	\label{threeterms2}
	\end{equation}
	where $A$ and $B$ are coefficients.
	
	\begin{figure}[h]
		\centering
		\includegraphics[width=0.5\linewidth]{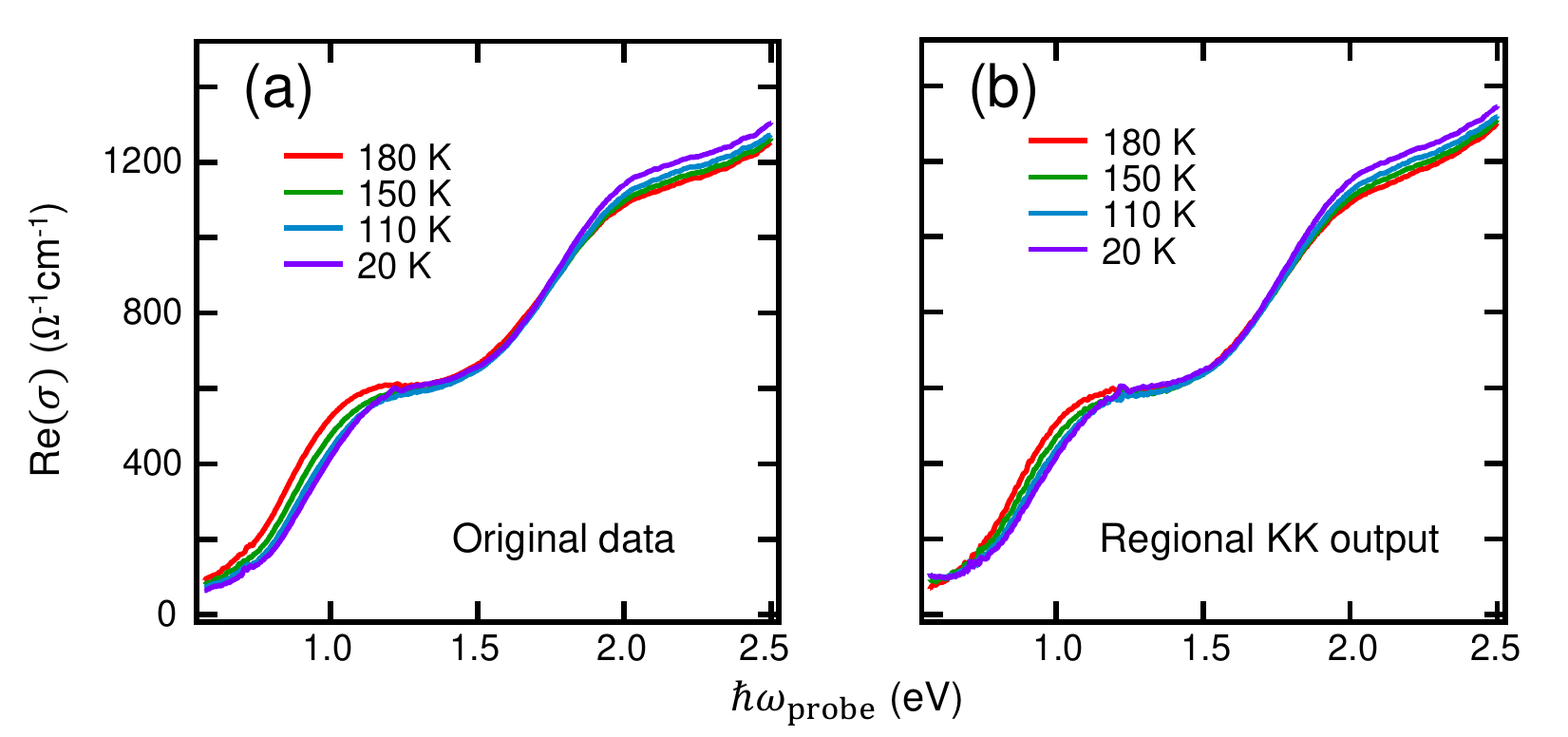}
		\caption{Benchmark test for the regional KK algorithm using the temperature dependent data set. (a)~Temperature dependent optical conductivity spectra directly from experiment used as the benchmark. (b)~Regional KK output, where the known 20~K conductivity spectrum is used to fit the $A$ and $B$ coefficients, and the reflectivity spectra for the rest of the temperatures are input into regional KK to output their respective conductivity spectra. Close agreement of conductivity spectra for all temperatures with (a) is observed.}
		\label{KKbenchmark}
	\end{figure}
	
	We fitted $A$ and $B$ from the known static $\sigma$ data at 20~K. In the nonequilibrium scenario, $\Delta R/R$ due to the optical pump will enter $\phi(\omega)$ to affect $\theta(\omega)$. $A$ and $B$ are also expected to change slightly due to nonzero $\Delta R/R$ in these unmeasured ranges. However, we found that when $\Delta R/R\ll1$, it is still numerically accurate to keep the nonequilibrium $A$ and $B$ constants to be the same as their equilibrium values, because the first and third terms in Eq.\,\ref{threeterms2} are off resonant in frequency.
	
	We did a benchmark test to prove the validity of this protocol. Figure\,\ref{KKbenchmark}(a) shows the static $\sigma$ at various temperatures. In Fig.\,\ref{KKbenchmark}(b), the 20~K curve is still the static one, while the 110~K, 150~K, 180~K curves are outputs from Eq.\,\ref{threeterms2}, where $A$ and $B$ coefficients are results from fitting to the 20~K data, and  the differences of reflectivity, $\Delta R_x = R_x-R_{20~\text{K}}$ ($x=$110~K, 150~K, 180~K), were input to the $\phi$ term. The close agreement between regional KK output at 110~K, 150~K, and 180~K in Fig.\,\ref{KKbenchmark}(b) and the experimental data in Fig.\,\ref{KKbenchmark}(a) suggests that the regional KK algorithm is accurate enough to give $\Delta\sigma$ when $\Delta R/R$ is small. None of our pump induced $\Delta R/R$ exceeds that induced by temperature (difference between 180~K and 20~K), and therefore, the method is expected to work well for our entire measurement.
	
	Empirically speaking, we found that the most critical factor impacting the robustness of the algorithm is the probe energy width of the experiment. For wider measurement ranges, the definite integral term for calculating the reflection phase (the middle term of Eq.\,\ref{threeterms}) becomes more dominant, and the algorithm appears more robust. This is because the equation used for fitting the reflection phase in Eq.\,\ref{threeterms2} contains two poles that are located exactly at the energy boundaries of the measurement. The KK transformed signals are inevitably subject to numerical artifacts at the poles and energies around the poles, but empirically we found that the artifact can be mitigated when the energy boundaries get further and further apart, that is, the range within which $\Delta R/R$ is experimentally measured gets wider. In our case, we did find artifacts associated with the poles (see the slight upturns of a few curves in Fig.\,\ref{KKbenchmark}(b) at the low-energy boundary for example), but our measurement range (0.55 eV to 2.2 eV) is large enough so that the artifacts are contained within a manageable extent.
	
	\subsection{B. Subtracting the differential optical conductivity}
	
	The regional KK transform outputs pump-induced differential conductivity spectra $\Delta\sigma$ for both the 0.3~eV pump ($\Delta\sigma_\text{0.3~eV}$) and the 1~eV pump ($\Delta\sigma_\text{1~eV}$) cases at various fluences and time delays. In Fig.~4 of the main text, we report analysis of difference spectra $\Delta(\Delta\sigma)=\Delta\sigma_\text{0.3~eV}-A\times\Delta\sigma_\text{1~eV}$, where $A$ is the scale factor, to account for the unique spectral signatures of the coherent non-thermal regime that are exclusively related to the subgap strong-field drive but not the photocarrier doping effect. Here, we describe how this subtraction was performed, and our way to determine the proper scaling factor multiplying $\Delta\sigma_\text{1~eV}$ in the subtraction.
	
	\begin{figure}[h]
		\centering
		\includegraphics[width=0.5\linewidth]{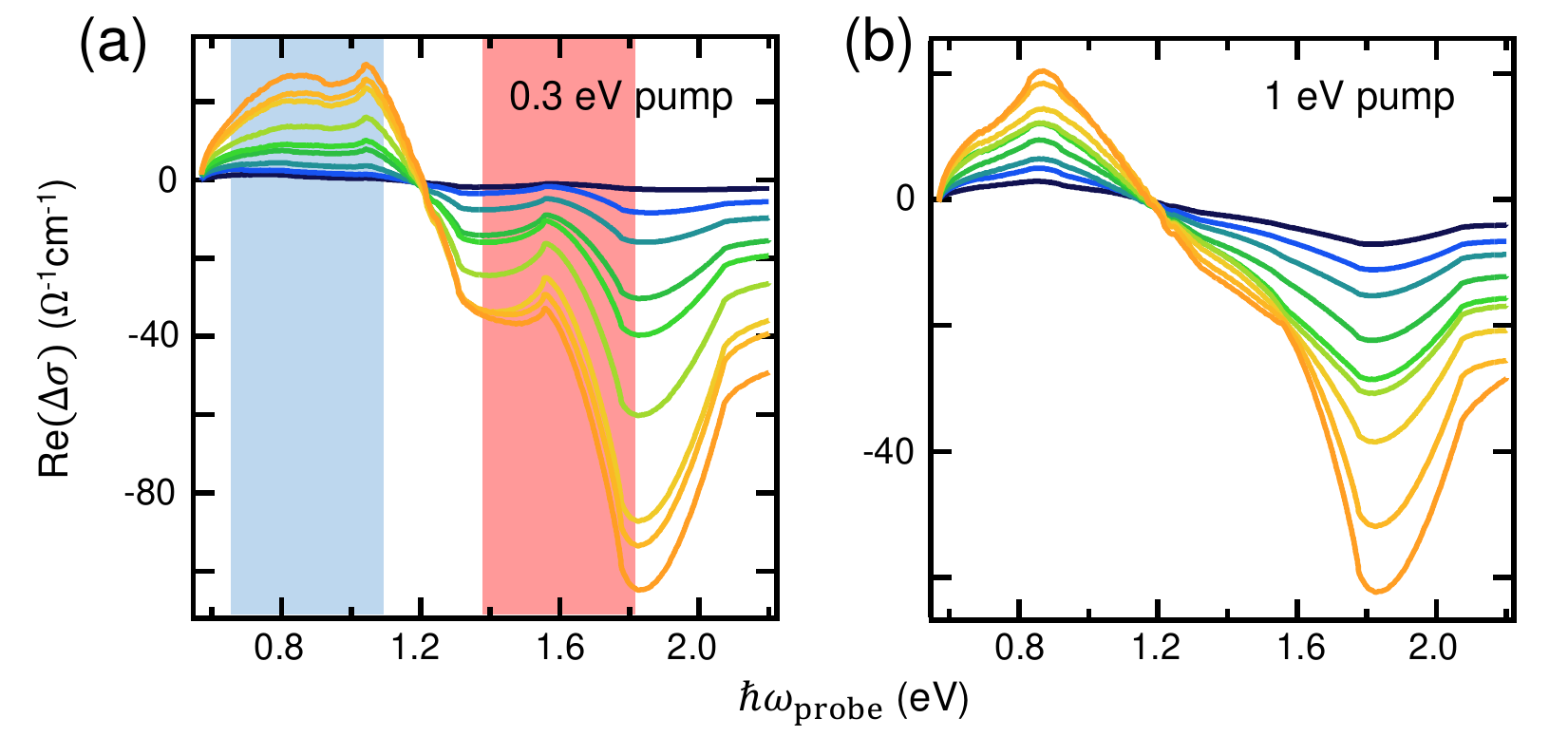}
		\caption{Comparison of $\Delta\sigma$ spectra for 0.3~eV pump (a) and 1~eV pump (b) at $t=0$~ps. Curves from blue to orange represent low to high fluences. 0.3~eV pump fluences range from 3~mJ/cm$^2$ to 30.4~mJ/cm$^2$. 1~eV pump fluences range from 0.57~mJ/cm$^2$ to 6.9~mJ/cm$^2$. Blue and red shaded regions in (a) highlight spectral ranges where additional modifications develop on the 0.3~eV pump data compared to the 1~eV pump data.}
		\label{dsigmacompare}
	\end{figure}
	
	Figures~\ref{dsigmacompare}(a) and (b) show a comparison between $\Delta\sigma_\text{0.3~eV}$ and $\Delta\sigma_\text{1~eV}$ across all fluences at time zero ($t=0$~ps). The probe energy ranges that show apparent modifications in $\Delta\sigma_\text{0.3~eV}$ data compared to $\Delta\sigma_\text{1~eV}$ are marked by the blue shade, where the positive peak looks flattened out, and the red shade, where a bump appears in the negative portion of the signal. In addition, a robust isosbestic point can be identified in both data sets at the same probe energy (1.2~eV) for all fluences. Generally speaking, for spectroscopic studies, an isosbestic point represents a frequency where measurement is most accurate, and is usually used as a reference point \cite{Eschenmoser1977SM}. In our case, the fact that it lies outside the blue and red shades (where spectral modifications obviously take place) strongly suggests that the probe energy of 1.2~eV, and energies that are right in the vicinity of it, are not influenced by the strong-field modification effect. Therefore, we chose the probe energy range between 1.1~eV to 1.3~eV as the reference, scaled $\Delta\sigma_\text{1~eV}$ to obtain the best matching with $\Delta\sigma_\text{0.3~eV}$ data in this range, and calculated $\Delta(\Delta\sigma)=\Delta\sigma_\text{0.3~eV}-A\times\Delta\sigma_\text{1~eV}$. This procedure was repeated for all time delays, producing the colormap in Fig.~4(a) of the main text.
	
	\begin{figure}[h]
		\centering
		\includegraphics[width=0.5\linewidth]{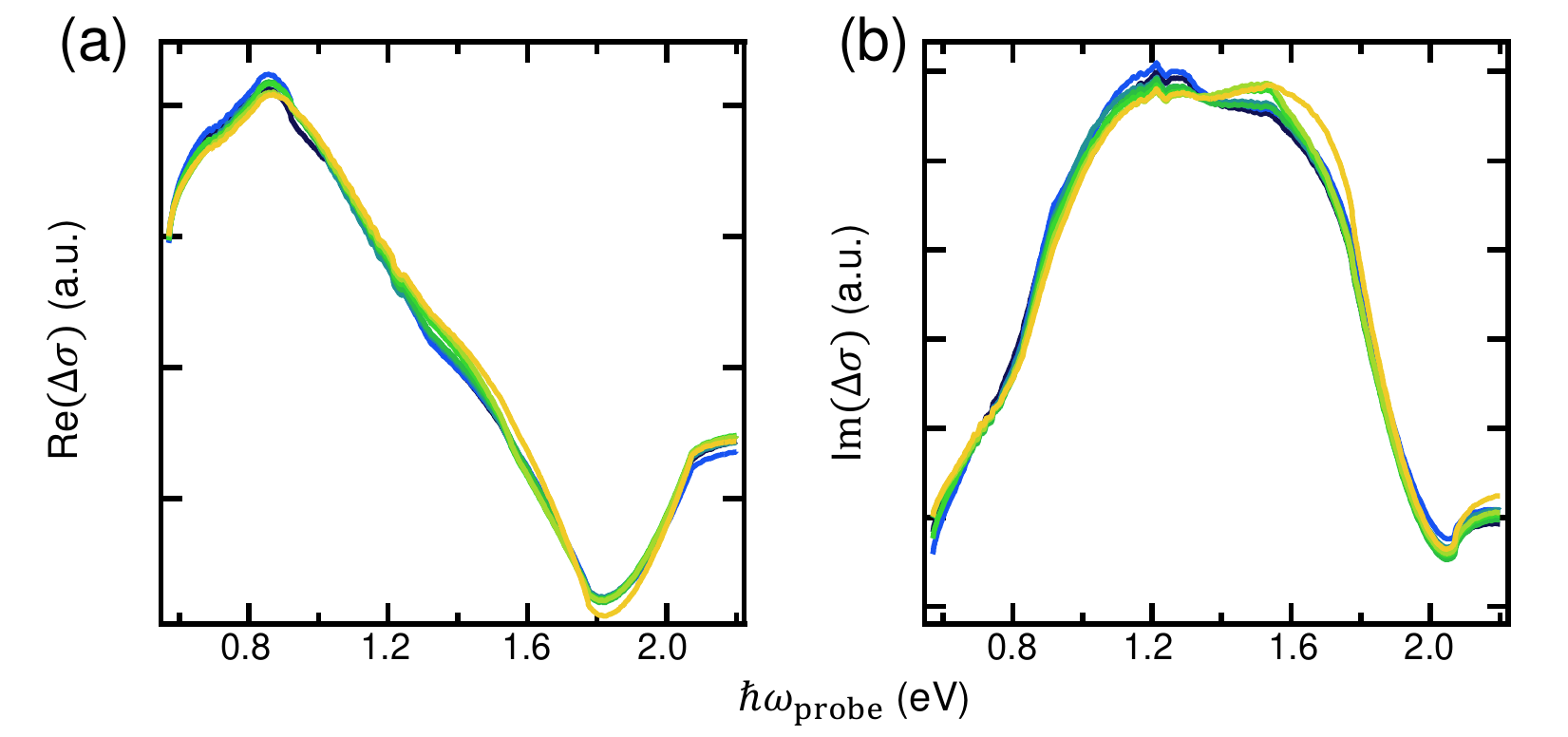}
		\caption{Scaling analysis for the real (a) and imaginary (b) parts of $\Delta\sigma$ for 1~eV pump at $t=0$~ps. Curves from blue to orange represent low to high fluences, and are multiplied by their respective scaling factors to make the traces overlap to the largest extent. Nice overlap is seen after scaling for both the real and the imaginary parts.}
		\label{scaling1eV}
	\end{figure}
	
	As shown in Fig.\,\ref{scaling1eV}, both the real and the imaginary parts of $\Delta\sigma_\text{1~eV}$ at $t=0$~ps scale well for all fluences, so the scaling factor in the $\Delta(\Delta\sigma)$ equation can simply account for the amplitude difference, and it is not important which fluence of $\Delta\sigma_\text{1~eV}$ is selected for the subtraction.

	\section{S6. Density functional theory simulations for bandwidth broadening}
	
	In Figs.\,4(d) and (e) of the main text, we report the expected change to optical conductivity by considering a bandwidth broadening process. The simulation outcomes were used to fit experimental $\Delta(\Delta\sigma)$ spectra to quantify the amount of bandwidth modification $(W-W_\text{eq})/W_\text{eq}$ as a function of fluence and time delay. Here we present details of the density functional theory (DFT) simulation and the way to fit data.
	
	\begin{figure}[h]
		\centering
		\includegraphics[width=0.6\linewidth]{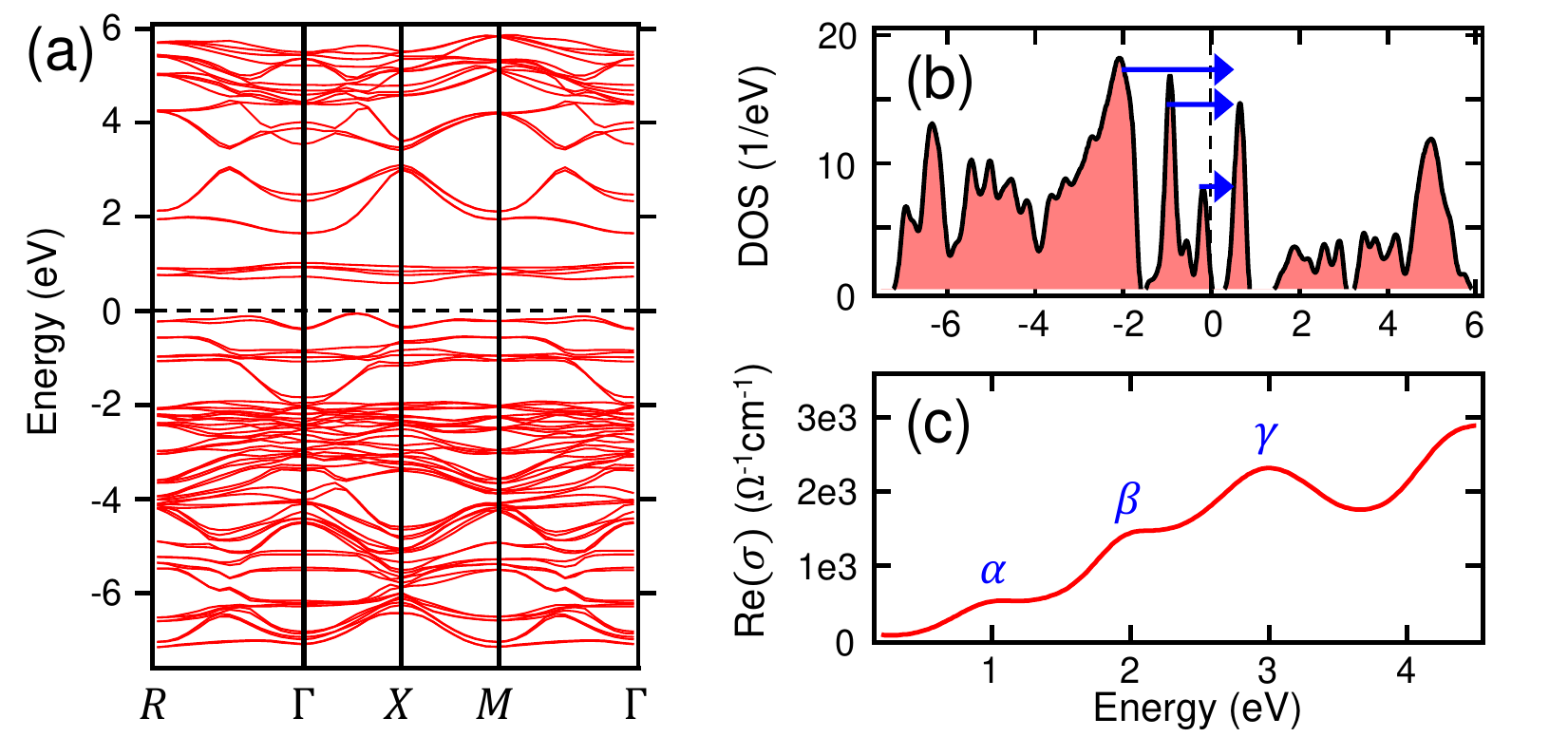}
		\caption{Static electronic properties of \CRO\ calculated by DFT. (a)~Band structure. (b)~Total density of states (DOS). Blue arrows indicate the $\alpha$, $\beta$, and $\gamma$ transition peaks in (c). (c)~Optical conductivity spectrum. Dashed lines show Fermi levels.}
		\label{DFTstatic}
	\end{figure}
	
	We used the structural parameters in Ref.\,\cite{Braden1998SM}, considered the collinear antiferromagnetic (AFM) structure along the $b$ axis, and applied the DFT$+U$ (static $U=3.5$~eV) method to simulate the static electronic properties of \CRO. Figures~\ref{DFTstatic}(a), (b), and (c) show the calculated band structure, total density of states (DOS), and conductivity spectrum, respectively. The Mott gap clearly opens up around the Fermi level when both the AFM structure and the Coulomb correlation are considered. Flat bands near the Fermi level are mostly contributed by $d$ orbitals of Ru, leading to concentrated DOS peaks. Three optical transitions across the DOS peaks clearly manifest in the optical conductivity spectrum as $\alpha$, $\beta$, and $\gamma$ transition peaks. This is in agreement with previous DFT and experimental studies on \CRO\ \cite{Fang2004SM,Jung2003SM}.
	
\begin{figure}[h]
		\centering
		\includegraphics[width=0.56\linewidth]{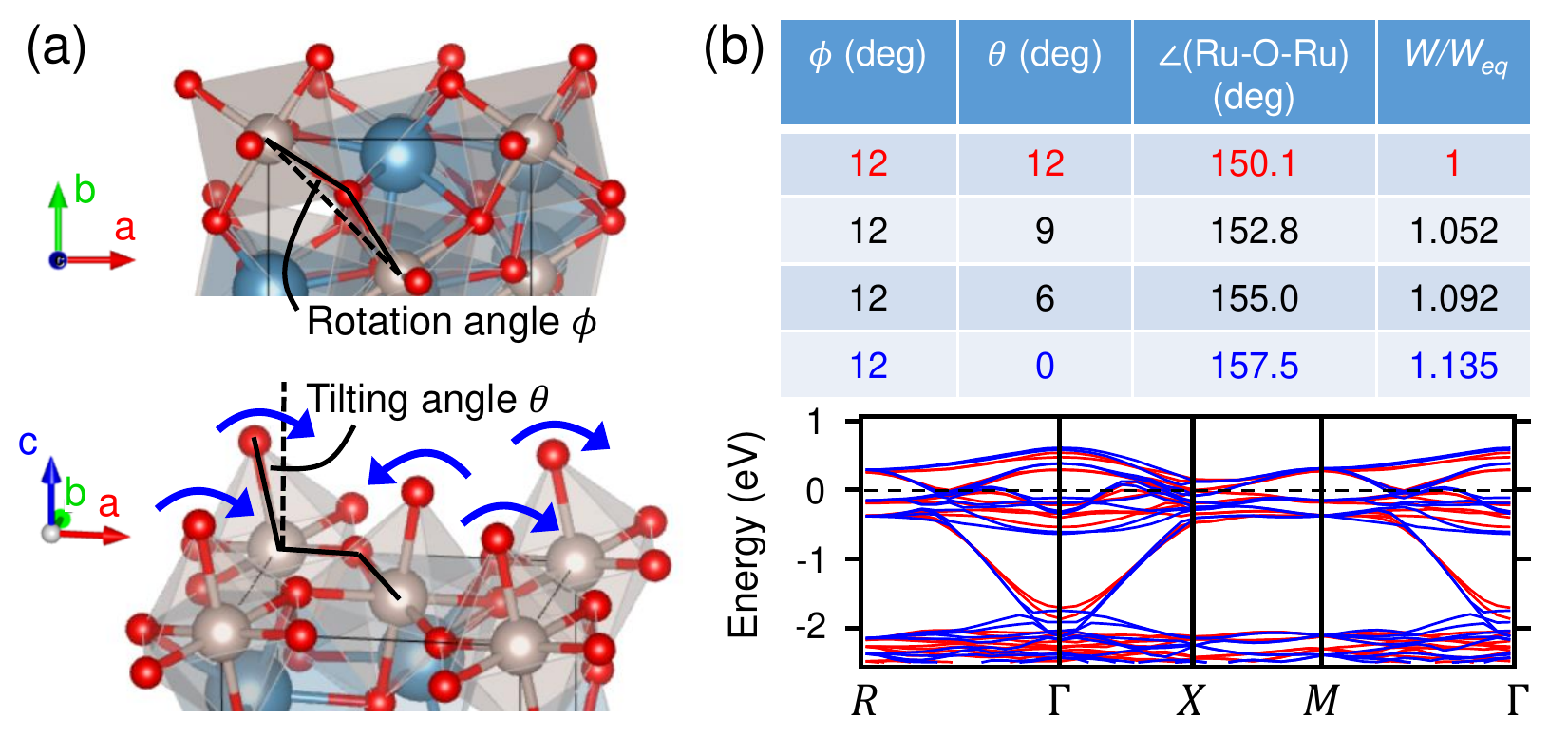}
		\caption{Method to simulate the effect of bandwidth broadening. (a)~There are two angles related to the lattice distortion in \CRO. Top: the rotation angle $\phi$ about the $c$~axis. Bottom: the tilting angle $\theta$ about an in-plane axis. (b)~Examples of how tuning $\theta$ affects the Ru-O-Ru bond angle and therefore, the bandwidth. Top: table displaying combinations of angles and the resulting modification to the bandwidth $W$ from the equilibrium $W_\text{eq}$. Bottom: band structures using the red and blue parameter conditions from the top table with a nonmagnetic structure and $U=0$~eV. Red (blue) bands correspond to the red (blue) parameter set.}
		\label{DFTbandwidth}
	\end{figure}

	To simulate the effect of bandwidth broadening, we changed the structural input parameters. In \CRO, each RuO$_6$ octahedron undergoes two types of distortions compared to the K$_2$NiO$_4$ structure (I4/mmm), leading to significant modifications to the in-plane hopping amplitudes, and therefore, the bandwidth. Figure~\ref{DFTbandwidth}(a) summarizes the two types of distortions, the rigid rotation of the octahedron around the $c$ axis by the angle $\phi$, and the rigid tilting of the octrahedron around an in-plane ($ab$ plane) axis by the angle $\theta$. In the static low temperature AFM state, $\phi=12^{\circ}$ and $\theta=12^{\circ}$, and the Ru-O-Ru bond angle $\angle(\text{Ru-O-Ru})=150.1^{\circ}$.
	
	We broadened the bandwidth $W$ in the simulation by reducing the tilting angle $\theta$ of the structure (blue arrows in Fig.\,\ref{DFTbandwidth}(a) bottom panel), while keeping all other structural parameters the same; this will make $\angle(\text{Ru-O-Ru})$ approach $180^{\circ}$, and broaden the bandwidth according to the empirical formula $W\propto [\text{cos}\angle(\text{Ru-O-Ru})]^2$ \cite{Woods2000SM}. It is worth noting that the logic of choosing $\theta$ to change while keeping $\phi$ a constant is based on the well known fact that $\theta$ responds much more sensitively to Sr doping \cite{Fang2004SM}, temperature \cite{Braden1998SM}, and applied current \cite{Bertinshaw2019SM} than $\phi$. In addition, the coherent $A_g$ phonon mode at 3.8~THz, which consists majorly of the tilting motion of RuO$_6$ octahedra, shows robust anomalies across the AFM ordering \cite{Lee2019SM} and orbital ordering \cite{Lee2018SM} temperatures. These all suggest that the tilting distortion is a crucial structural parameter in \CRO\ which responds sensitively to magnetic and electronic ground states. This justifies us adjusting $\theta$ for simulating the bandwidth renormalization induced by the strong-field drive, even though the drive does not directly modify $\theta$. The table in Fig.\,\ref{DFTbandwidth}(b) shows examples of combinations of structure parameters, and the resulting ratio of the modified bandwidth to the static equilibrium bandwidth, $W/W_\text{eq}$, estimated from $W\propto [\text{cos}\angle(\text{Ru-O-Ru})]^2$. To make sure that $W$ is actually modified, we simulated the nonmagnetic crystal with $U=0$~eV using the red and blue parameter sets in the table; the calculated bands are shown in the bottom panel of Fig.\,\ref{DFTbandwidth}(b). The nonmagnetic setting with $U=0$~eV fully collapses the Mott gap, making it easier for us to identify a bandwidth change. As clearly observed in the bottom panel of Fig.\,\ref{DFTbandwidth}(b), the blue parameter set indeed leads to a broadened bandwidth compared to the red parameter set.
	
	\begin{figure}[h]
		\centering
		\includegraphics[width=0.8\linewidth]{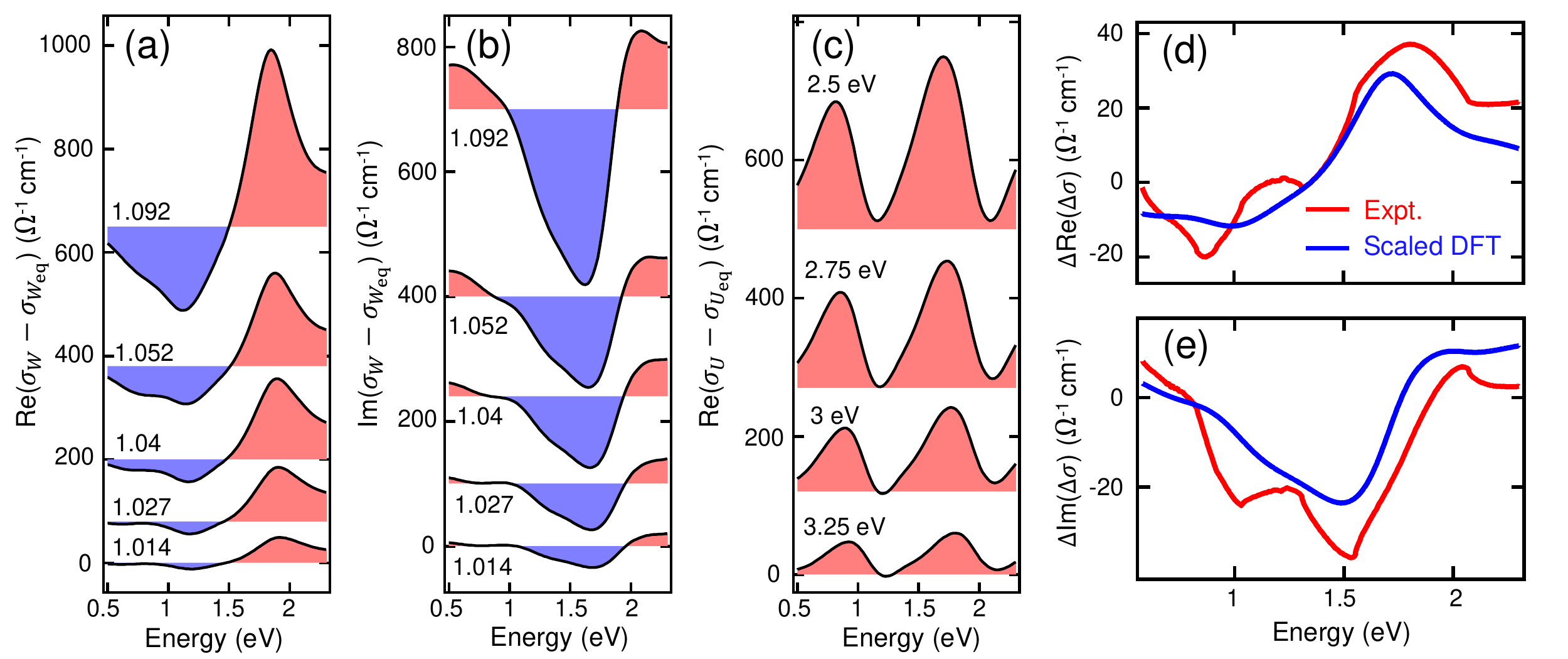}
		\caption{DFT simulation outcomes of bandwidth modification. Calculated change in the real (a) and imaginary (b) parts of conductivity due to bandwidth broadening. Bandwidth narrowing would give spectra off from these traces by a minus sign. $W/W_\text{eq}$ values are labeled. (c)~Calculated change in conductivity due to dynamical $U$ modification. Values of $U$ are labeled. (d) and (e) show an example of scaling the simulated DFT spectra to fit the experimental $\Delta(\Delta\sigma)$ data to quantify $W/W_\text{eq}$. Real and imaginary parts are scaled by a common factor. Curves are offset for clarity from (a) to (c).}
		\label{DFTtU}
	\end{figure}
	
	Figures~\ref{DFTtU}(a) and (b) show calculated modifications to conductivity by changing the bandwidth by various amounts; $W/W_\text{eq}$ are labeled on each curve. Both the real and the imaginary part show agreement with the experimental $\Delta(\Delta\sigma)=\Delta\sigma_\text{0.3~eV}-A\times\Delta\sigma_\text{1~eV}$. In contrast, if we consider another scenario where modification to Coulomb correlation $U$ occurs \cite{TancogneDejean2018SM}, the change in conductivity would look very different, and would not match $\Delta(\Delta\sigma)$; see Fig.\,\ref{DFTtU}(c). Finally, given the simulation results of bandwidth modification, the method we used to quantitatively determine experimental $W/W_\text{eq}$ is scaling the $W/W_\text{eq}=1.052$ curve (Since the difference in $\sigma$ roughly grows in proportion with $(W-W_\text{eq})/W_\text{eq}$, it does not matter which curve to pick here.) in Figs.\,\ref{DFTtU}(a) and (b) by a common factor to fit experimental data, as shown in Figs.\,\ref{DFTtU}(d) and (e). The same factor is then multiplied to the $(W-W_\text{eq})/W_\text{eq}$ ratio set for the simulation to give the actual experimental $(W-W_\text{eq})/W_\text{eq}$, assuming linear proportionality when the fractional modification is small. Error bars in the main text are quantified by the standard deviation between the calculation and experiment.

	\section{S7. Floquet calculation of bandwidth renormalization}
	
	In the main text, we discussed that the ultrafast bandwidth renormalization (UBR) observed in the subgap strong-field pump data at exactly time zero originates from a Floquet engineering mechanism. To give a quantitative estimate of the UBR due to the Floquet mechanism and compare with our experiment, we followed Ref.\,\cite{Mentink2015SM} and used the Floquet-driven two-site cluster model therein. The two-site cluster model takes the periodic-field-dependent electronic hopping into account, but significantly simplifies the problem. Dynamical mean-field theory for systems with extended dimensions also show good agreement with the two-site model. According to Ref.\,\cite{Mentink2015SM}, when the Mott insulator is strongly coupled ($U\gg t$), the ratio between the light-modified bandwidth and the static bandwidth is
	\begin{equation}
	\frac{W}{W_\text{eq}}=\sqrt{\sum_{n=-\infty}^{\infty}\frac{J_\abs{n}(\mathcal{E})^2}{1+n\omega/U}},
	\label{Floquet}
	\end{equation}
	where $\mathcal{E}=eaE_0/(\hbar\omega)$ is the Floquet parameter, $a$ is the lattice constant, $E_0$ is the field amplitude, $\omega$ is the pump frequency, and $J_n(x)$ is the $n$th Bessel function.
	
	We input our experimental pumping conditions into the equation, $a=5.6$~\AA, and a range of $U$ from 3~eV to 3.5~eV, with no other adjustable parameter. Since hopping $t=0.23$~eV, the $t/U\ll1$ condition holds. The result of this calculation using Eq.\,\ref{Floquet} is reported in Fig.\,4(e) of the main text.

	\section{S8. Relation between differential reflectivity and d-h pair density}
	
	In this section, we present additional clarifications of the relation between the differential reflectivity and the d-h pair density. Two specific problems will be addressed. One is the proof of proportionality between differential reflectivity and the pair generation rate. We will then expand  the model to take into account pair distribution functions, and show that the pair distribution function must undergo a crossover in the fluence scaling whenever the differential reflectivity spectra at different fluences cannot scale.
	
	First, we identify that any pump-induced spectral modification at 0.1~ps (time delay at which $\Delta R/R$ peaks) should originate from the photo-excited pairs. The coherent Floquet modification and heating is expected to provide negligible contribution since these processes are separated in time from 0.1~ps. For photo-excited d-h pairs with a density of $n$, their impact on the reflectivity spectrum can be expanded as
	\begin{equation}
	\Delta R(\omega)=R(n,\omega)-R(n=0,\omega)\approx n\left.\frac{\partial R(n,\omega)}{\partial n}\right\rvert_{n=0}
	\label{simpleTaylor}
	\end{equation}
	where we retain only the linear term in the Taylor series (given the condition of $\Delta R/R\ll1$, which holds for our entire fluence range), and assume that the coefficient $\left.\frac{\partial R(n,\omega)}{\partial n}\right\rvert_{n=0}$ is nonzero in general. Note that the expression remains valid for all types of photo-induced spectral modifications, including peak shift and broadening, and the resulting relation of $\Delta R(\omega)\propto n$  has been frequently used by the ultrafast optical community to describe quasiparticle dynamics in various photo-excited gapped systems \cite{Gedik2004SM,Chia2006SM,Demsar1999SM}.
	Assuming a simplified scenario where pair generation is uniform in rate within the pump pulse duration $\Delta t$, one can write the rate $\Gamma=n/\Delta t$. This relation, combined with Eq.\,\ref{simpleTaylor}, establishes $\Delta R(\omega)\propto n\propto\Gamma$. The reason we leave $\Gamma$ in arbitrary units is because the coefficient relating $\Delta R(\omega)$ with $\Gamma$ is not determined quantitatively. But establishing the proportionality is sufficient for us to perform the scaling analysis in this work.
	
\begin{figure}[h]
		\centering
		\includegraphics[width=0.35\linewidth]{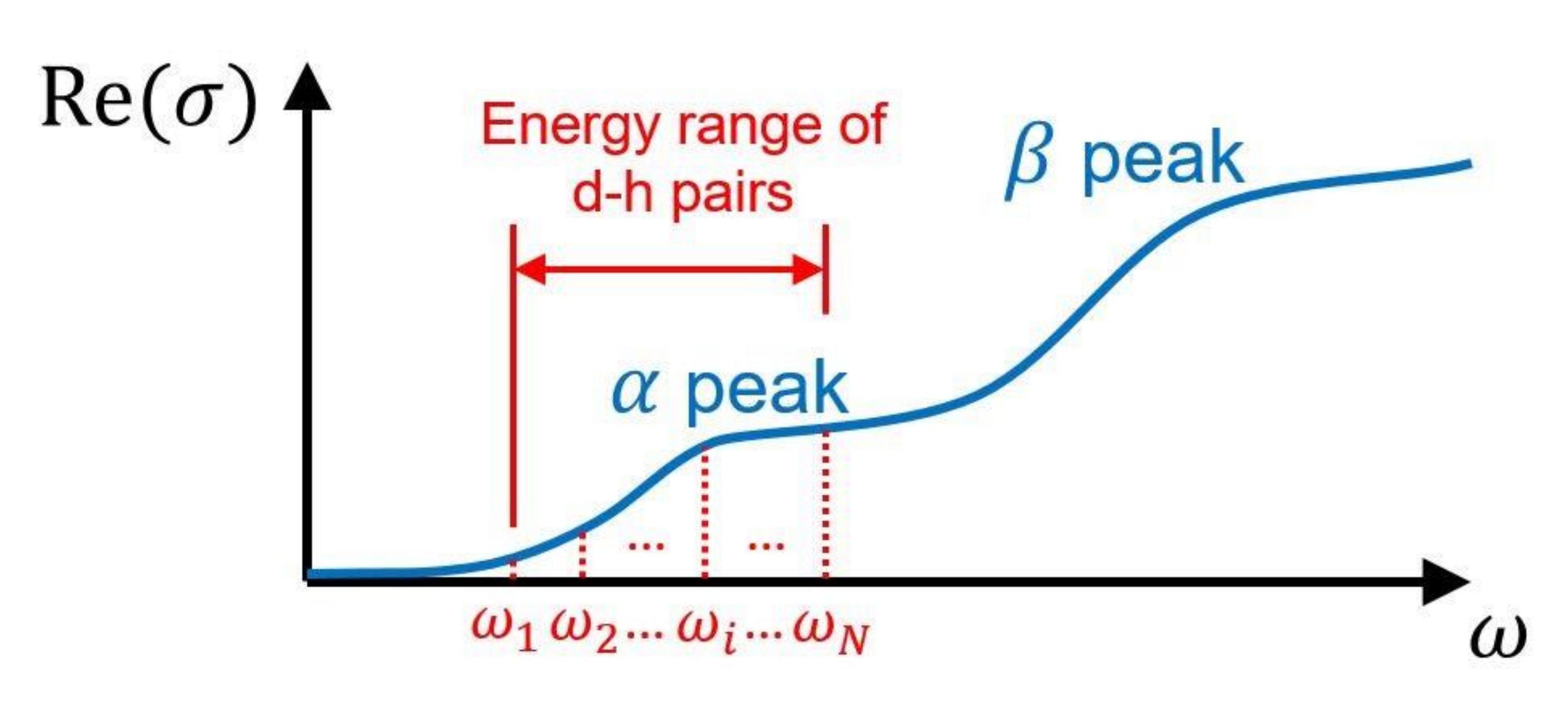}
		\caption{Schematic showing the d-h pair population on the optical conductivity (joint density of states) spectrum that can be distinguished by the pair energy.}
		\label{Rnmodel}
	\end{figure}

	We then consider an expanded model where $\Delta R$ is influenced by, not one, but multiple species of d-h pairs distinguished by the pair energies. 
	For the nonequilibrium situation where photo-excited d-h pairs are occupying the upper and lower Hubbard bands, the pairs can be labeled in the joint density of states spectrum by their energies (doublon and holon energies combined for each pair) $\omega_i$. If we represent the number of pairs with energy $\omega_i$ as $n_{\omega_i}$, and divide the energy window within which pairs populate into a total of $N$ bins, the pair distribution function can then be represented by a collection of $\{n_{\omega_i}\}$ for  $i\in\{1,2,...,N\}$; see Fig.\,\ref{Rnmodel} for a schematic.
	
	We are interested in finding the influence of $\{n_{\omega_i}\}$ on the reflectivity spectrum. Consider the general case where pairs with different energies impact the spectrum differently, the photo-carrier induced reflectivity change can be expanded into the Taylor series as
	\begin{align}
		\Delta R(F,\omega) & =R(n_{\omega_1},n_{\omega_2},...,n_{\omega_N},\omega)-R(n_{\omega_1}=n_{\omega_2}=...=n_{\omega_N}=0,\omega)\\
		& \approx \sum_{i=1}^{N}n_{\omega_i}(F)\left.\frac{\partial R(n_{\omega_1},n_{\omega_2},...,n_{\omega_N},\omega)}{\partial n_{\omega_i}}\right\rvert_{n_{\omega_1}=n_{\omega_2}=...=n_{\omega_N}=0}
		\label{cmplxTaylor}
	\end{align}
	where again only the linear terms are retained, and $n_{\omega_i}$ depends on pump fluence $F$ as $n_{\omega_i}(F)$.
	For two fluence values $F_1$ and $F_2$, the ratio of the reflectivity spectra
	\begin{equation}
	\frac{\Delta R(F_1,\omega)}{\Delta R(F_2,\omega)}=\frac{\sum_{i=1}^{N}n_{\omega_i}(F_1)\left.\frac{\partial R(n_{\omega_1},n_{\omega_2},...,n_{\omega_N},\omega)}{\partial n_{\omega_i}}\right\rvert_{n_{\omega_1}=n_{\omega_2}=...=n_{\omega_N}=0}}{\sum_{i=1}^{N}n_{\omega_i}(F_2)\left.\frac{\partial R(n_{\omega_1},n_{\omega_2},...,n_{\omega_N},\omega)}{\partial n_{\omega_i}}\right\rvert_{n_{\omega_1}=n_{\omega_2}=...=n_{\omega_N}=0}}
	\label{dRratio}
	\end{equation}
	should be  $\omega$-dependent in general. However, in certain regimes of photo-excitation, the pair distribution follows well-defined scaling functions, that is, for $\forall i,j\in\{1,2,...N\}$, there always exists a constant $C$, that makes $C\cdot n_{\omega_i}(F)/n_{\omega_j}(F)=1$. This is equivalent to writing $n_{\omega_i}(F)=C_if(F)$, where $C_i$ is  $F$-independent and $f(F)$ is a universal scaling function.
	
	We give three concrete cases where such scaling functions exist: \\
	(1) For above-gap photo-doping pump, $f(F)=F$.\\
	(2) Within the deep multi-photon regime \cite{Oka2012SM} ($\gamma_K\gg1$), $f(F)=F^{a/2}$ ($a>2$).\\
	(3) Within the deep tunneling regime \cite{Oka2012SM} ($\gamma_K\ll1$), $f(F)=e^{-b/\sqrt{F}}$.
	
	The fluence dependence can then be factored out as
	\begin{equation}
	\Delta R(F,\omega)=f(F)\sum_{i=1}^{N}C_i\cdot\left.\frac{\partial R(n_{\omega_1},n_{\omega_2},...,n_{\omega_N},\omega)}{\partial n_{\omega_i}}\right\rvert_{n_{\omega_1}=n_{\omega_2}=...=n_{\omega_N}=0}
	\label{factoreddRratio}
	\end{equation}
	so that the ratio $\frac{\Delta R(F_1,\omega)}{\Delta R(F_2,\omega)}=\frac{f(F_1)}{f(F_2)}$  becomes $\omega$-independent. The $\Delta R(\omega)$ spectrum therefore “scales” for various pump fluences, and we refer to this scenario as successful scaling. For the specific case of insulating systems, photo-excitation typically causes spectral weight transfers, which manifest as zero crossing features in  $\Delta R(\omega)$. For this type of spectra, a successful scaling ensures that the zero-crossing energy does not shift with fluence (as observed in Fig. 2(h) of the main text), leading to an isosbestic point in the reflectivity spectrum, $R(\omega)=\Delta R(\omega)+R_\text{eq}(\omega)$, where $R_\text{eq}(\omega)$ represents the spectrum in equilibrium.
	
	On the other hand, according to Eq. \ref{dRratio}, unsuccessful scaling, defined as $\frac{\Delta R(F_1,\omega)}{\Delta R(F_2,\omega)}$ being  $\omega$-dependent, occurs for the Keldysh crossover \cite{Oka2012SM} during which there is no universal scaling function $f(F)$ that can be factored out from $n_{\omega_{i}}$; pair distribution change during the Keldysh crossover causes $n_{\omega_{i}}$ at different $\omega_{i}$ to scale differently with $F$. Absence of an isosbestic point in $R(\omega)=\Delta R(\omega)+R_\text{eq}(\omega)$, which is equivalent to the statement that the zero-crossing energy in $\Delta R(\omega)$ shifts with fluence, should be a manifestation of unsuccessful scaling, and therefore, can serve as evidence for the Keldysh crossover.

	\section{S9. Lorentz model fitting of transient conductivity}
	
	Here we examine if the UBR due to Floquet engineering can be directly identified from the conductivity spectra. The idea is to fit $\alpha$ and $\beta$ peaks with Lorentzians, and see if UBR manifests in their peak widths as a function of pump electric field strength $E_\text{pump}$.
	
\begin{figure}[h]
		\centering
		\includegraphics[width=0.6\linewidth]{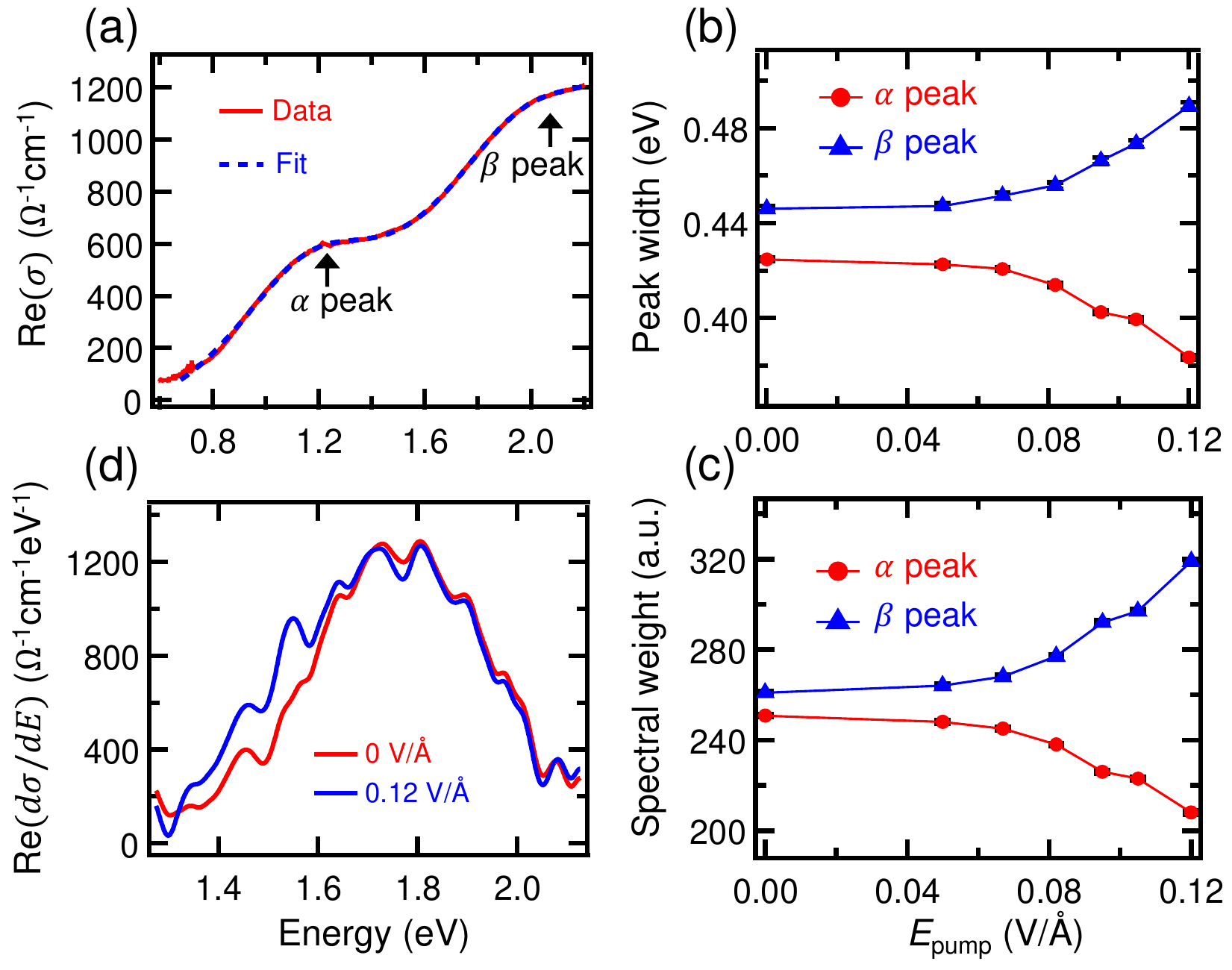}
		\caption{Lorentz model fitting of optical conductivity. (a) Equilibrium conductivity at 20 K fitted by Eq.\,\ref{Loretzfiteq}. (b) Peak widths and (c) amplitudes for the $\alpha$ and $\beta$ bands versus pump field. (d) First order derivative of conductivity to show earlier onset of $\beta$ peak conductivity in the laser-driven sample (blue) than the sample in equilibrium (red).}
		\label{Lorentzfitting}
	\end{figure}

	We set up a fitting equation that expresses conductivity $\sigma$ versus probe photon energy $E$ as
	\begin{equation}
	\text{Re}(\sigma)=p_0+ p_1 E+p_2 E^2+\frac{A_\alpha\Delta E_\alpha}{(E-E_\alpha)^2+(\Delta E_\alpha)^2}+\frac{A_\beta\Delta E_\beta}{(E-E_\beta)^2+(\Delta E_\beta)^2},
	\label{Loretzfiteq}
	\end{equation}
	which contains polynomial terms up to quadratic order to account for the background spectral weight, and two Lorentzians to account for the $\alpha$ and $\beta$ peaks. $A_\alpha$ ($A_\beta$), $E_\alpha$ ($E_\beta$), $\Delta E_\alpha$ ($\Delta E_\beta$) represent spectral weight, center energy, and peak width of the $\alpha$ ($\beta$) peak, respectively. Figure \ref{Lorentzfitting}(a) shows the agreement between the fit and the equilibrium conductivity spectrum using Eq.\,\ref{Loretzfiteq}.
	
	The similar fitting procedure is carried out for nonequilibrium $\text{Re}(\sigma)$ at various $E_\text{pump}$ values. In order to only take the Floquet effect into account, we express $\sigma = \sigma_\text{eq}+\Delta(\Delta\sigma(t=0))$, where $\sigma_\text{eq}$ is the equilibrium conductivity, and $\Delta(\Delta\sigma(t=0))$ is the time-zero nonthermal signal identified using the subtraction process (explained in Section S5B). Figure\,\ref{Lorentzfitting}(b) shows the extracted peak widths versus $E_\text{pump}$. Although the $\alpha$ peak narrows with $E_\text{pump}$, the $\beta$ peak clearly shows a broadening with $E_\text{pump}$ whose trend matches closely with that in Fig.\,4(e) of the main text. In addition, the $\alpha$ peak spectral weight transfers to the $\beta$ peak with increasing $E_\text{pump}$ (Fig.\,\ref{Lorentzfitting}(c)). The $\beta$ peak broadening can be directly identified in the conductivity spectra by performing an energy derivative; see Fig.\,\ref{Lorentzfitting}(d) for a comparison of $d\sigma/dE$ between the equilibrium and the laser-driven scenarios. The fact that the driven scenario shows an earlier onset of the beta peak in the 1.4~eV - 1.8~eV range corroborates the conclusion from our fitting.
	
	The broadening of the $\beta$ peak suggests a bandwidth increase of the $d_{xz}$ and $d_{yz}$ orbitals, which agrees with the conclusion from the DFT simulations reported in the main text. The observation of the $\alpha$ peak narrowing, however, requires more interpretations by future work. At a qualitative level, the $\alpha$ peak is expected to closely correlate with the hole population within the $d_{xy}$ orbital (which arises from orbital mixing of $d_{xy}$ into $d_{xz/yz}$ due to crystal field distortions), the appearance of the peak can thus respond sensitively to conditions other than a pure $d_{xy}$ bandwidth broadening effect. Indeed, the spectral weight decrease of the $\alpha$ peak suggests a decrease of the $d_{xy}$ hole population versus $E_\text{pump}$, which should be expected when DFT simulates a less distorted, bandwidth-broadened crystal by ``straightening" the Ru-O-Ru bonds \cite{Jung2003SM}. Therefore, a narrowing of the $\alpha$ peak can still be consistent with bandwidth broadening provided microscopic details are fully considered as in our first-principles calculations.


\end{document}